\documentclass[twocolumn,twocolappendix]{aastex63}

\usepackage{romannum}
\AtBeginDocument{\pagenumbering{arabic}}

\newcommand{\ha}{H$\alpha$}
\newcommand{\betauv}{$\beta_{\rm UV}$}
\newcommand{\betaopt}{$\beta_{\rm opt}$}
\newcommand{\Lbol}{$L_{\rm bol}$}
\newcommand{\MBH}{$M_{\rm BH}$}

\newcommand{\xj}[1]{{#1}} 
\newcommand{\xlin}[1]{{#1}}

\received{XXX}
\revised{YYY}
\accepted{ZZZ}

\submitjournal{AJ}

\shorttitle{ASPIRE broad \ha\ emitters}
\shortauthors{Lin et al.}

\begin{document}

\title{A SPectroscopic survey of biased halos In the Reionization Era (ASPIRE): Broad-line AGN at $z=4-5$ revealed by JWST/NIRCam WFSS }

\author[0000-0001-6052-4234]{Xiaojing Lin}
\affiliation{Department of Astronomy, Tsinghua University, Beijing 100084, China}
\email{linxj21@mails.tsinghua.edu.cn}
\affiliation{Steward Observatory, University of Arizona, 933 N Cherry Ave, Tucson, AZ 85721, USA}

\author[0000-0002-7633-431X]{Feige Wang}
\affiliation{Steward Observatory, University of Arizona, 933 N Cherry Ave, Tucson, AZ 85721, USA}

\author[0000-0003-3310-0131]{Xiaohui Fan}
\affiliation{Steward Observatory, University of Arizona, 933 N Cherry Ave, Tucson, AZ 85721, USA}

\author[0000-0001-8467-6478]{Zheng Cai}
\affiliation{Department of Astronomy, Tsinghua University, Beijing 100084, China}

\author[0000-0002-6184-9097]{Jaclyn B. Champagne}
\affiliation{Steward Observatory, University of Arizona, 933 N Cherry Ave, Tucson, AZ 85721, USA}

\author[0000-0002-4622-6617]{Fengwu Sun}
\affiliation{Steward Observatory, University of Arizona, 933 N Cherry Ave, Tucson, AZ 85721, USA}
\affiliation{Center for Astrophysics $|$ Harvard \& Smithsonian, 60 Garden St., Cambridge MA 02138 USA}

\author[0000-0002-3216-1322]{Marta Volonteri}
\affiliation{Institut d'Astrophysique de Paris, Sorbonne Universit\'e, CNRS, UMR 7095, 98 bis bd Arago, 75014 Paris, France}

\author[0000-0001-5287-4242]{Jinyi Yang}
\affiliation{Steward Observatory, University of Arizona, 933 N Cherry Ave, Tucson, AZ 85721, USA}

\author[0000-0002-7054-4332]{Joseph F. Hennawi}
\affiliation{Department of Physics, Broida Hall, University of California, Santa Barbara, CA 93106-9530, USA}
\affiliation{Leiden Observatory, Leiden University, P.O. Box 9513, NL-2300 RA Leiden, The Netherlands}

\author[0000-0002-2931-7824]{Eduardo Ba{\~n}ados}
\affiliation{Max Planck Institut f\"ur Astronomie, K\"onigstuhl 17, D-69117, Heidelberg, Germany}

\author[0000-0002-3026-0562]{Aaron Barth}
\affiliation{Department of Physics and Astronomy, 4129 Frederick Reines Hall, University of California, Irvine, CA 92697-4575, USA}

\author[0000-0003-2895-6218]{Anna-Christina Eilers}
\affiliation{MIT Kavli Institute for Astrophysics and Space Research, Massachusetts Institute of Technology, Cambridge, MA 02139, USA}

\author{Emanuele Paolo Farina}
\affiliation{Gemini Observatory, NSF's NOIRLab, 670 North A'ohoku Place, Hilo, HI 96720, USA}

\author[0000-0003-3762-7344]{Weizhe Liu}
\affiliation{Steward Observatory, University of Arizona, 933 N Cherry Ave, Tucson, AZ 85721, USA}

\author{Xiangyu Jin}
\affiliation{Steward Observatory, University of Arizona, 933 N Cherry Ave, Tucson, AZ 85721, USA}

\author[0000-0003-1470-5901]{Hyunsung D. Jun}
\affiliation{Department of Physics, Northwestern College, 101 7th St SW, Orange City, IA 51041, USA
}

\author[0000-0001-6106-7821]{Alessandro Lupi}
\affiliation{Dipartimento di Scienza e Alta Tecnologia, Universit\`a degli Studi dell'Insubria, via Valleggio 11, I-22100, Como, Italy}
\affiliation{INFN, Sezione di Milano-Bicocca, Piazza della Scienza 3, I-20126 Milano, Italy}
\affiliation{Dipartimento di Fisica ``G. Occhialini'', Universit\`a degli Studi di Milano-Bicocca, Piazza della Scienza 3, I-20126 Milano, Italy}

\author{Koki Kakiichi}
\affiliation{Cosmic Dawn Center (DAWN), Denmark}
\affiliation{Niels Bohr Institute, University of Copenhagen, Jagtvej 128, 2200 Copenhagen N, Denmark}

\author[0000-0002-5941-5214]{Chiara Mazzucchelli}
\affiliation{Instituto de Estudios Astrof\'{\i}sicos, Facultad de Ingenier\'{\i}a y Ciencias, Universidad Diego Portales, Avenida Ejercito Libertador 441, Santiago, Chile.}

\author[0000-0003-2984-6803]{Masafusa Onoue}
\affiliation{Kavli Institute for the Physics and Mathematics of the
Universe (Kavli IPMU, WPI), The University of Tokyo Institutes for Advanced Study, The University of Tokyo, Kashiwa, Chiba 277-8583, Japan}
\affiliation{Center for Data-Driven Discovery, Kavli IPMU (WPI), UTIAS, The University of Tokyo, Kashiwa, Chiba 277-8583, Japan}
\affiliation{Kavli Institute for Astronomy and Astrophysics, Peking
University, Beijing 100871, P.R.China}

\author{Zhiwei Pan}
\affiliation{Kavli Institute for Astronomy and Astrophysics, Peking University, Beijing 100871, People's Republic of China}
\affiliation{Department of Astronomy, School of Physics, Peking University, Beijing 100871, People's Republic of China}

\author{Elia Pizzati}
\affiliation{Leiden Observatory, Leiden University, P.O. Box 9513, 2300 RA Leiden, The Netherlands}

\author[0000-0003-2349-9310]{Sof\'ia Rojas-Ruiz}
\affiliation{Department of Physics and Astronomy, University of California, Los Angeles, 430 Portola Plaza, Los Angeles, CA 90095, USA}

\author[0000-0002-4544-8242]{Jan-Torge Schindler}
\affiliation{Hamburger Sternwarte, Universität Hamburg, Gojenbergsweg 112, D-21029 Hamburg, Germany}

\author{Benny Trakhtenbrot}
\affiliation{School of Physics and Astronomy, Tel Aviv University, Tel Aviv 69978, Israel}

\author[0000-0003-1659-7035]{Yue Shen}
\affiliation{Department of Astronomy, University of Illinois at Urbana-Champaign, Urbana, IL 61801, USA}
\affiliation{National Center for Supercomputing Applications, University of Illinois at Urbana-Champaign, Urbana, IL 61801, USA}

\author{Maxime Trebitsch}
\affiliation{Kapteyn Astronomical Institute, University of Groningen, 9700 AV Groningen, The Netherlands}

\author[0000-0001-5105-2837]{Ming-Yang Zhuang}
\affiliation{Department of Astronomy, University of Illinois Urbana-Champaign, Urbana, IL 61801, USA}


\author{Ryan Endsley}
\affiliation{Department of Astronomy, University of Texas, Austin, TX 78712, USA}

\author{Romain A. Meyer}
\affiliation{Max Planck Institut f\"ur Astronomie, K\"onigstuhl 17, D-69117, Heidelberg, Germany}

\author{Zihao Li}
\affiliation{Department of Astronomy, Tsinghua University, Beijing 100084, China}

\author{Mingyu Li}
\affiliation{Department of Astronomy, Tsinghua University, Beijing 100084, China}

\author{Maria Pudoka}
\affiliation{Steward Observatory, University of Arizona, 933 N Cherry Ave, Tucson, AZ 85721, USA}

\author{Wei Leong Tee}
\affiliation{Steward Observatory, University of Arizona, 933 N Cherry Ave, Tucson, AZ 85721, USA}

\author{Yunjing Wu}
\affiliation{Department of Astronomy, Tsinghua University, Beijing 100084, China}

\author{Haowen Zhang}
\affiliation{Steward Observatory, University of Arizona, 933 N Cherry Ave, Tucson, AZ 85721, USA}


 
\correspondingauthor{Xiaojing Lin}

\begin{abstract}

Low-luminosity AGNs with low-mass black holes (BHs) in the early universe are fundamental to understanding the BH growth and their co-evolution with the host galaxies.  Utilizing JWST NIRCam Wide Field Slitless Spectroscopy (WFSS), we perform a systematic search for broad-line \ha\ emitters (BHAEs) at $z\approx 4-5$ in 25 fields of the ASPIRE (A SPectroscopic survey of biased halos In the Reionization Era) project, covering a total area of 275 arcmin$^2$. We identify 16 BHAEs with FWHM of the broad components spanning from $\sim$ 1000 km s$^{-1}$ to  3000 km s$^{-1}$. Assuming the broad linewidths arise due to Doppler broadening around BHs, the implied BH masses range from $10^7$ to $10^{8}~M_\odot$, with broad \ha-converted bolometric luminosity of $10^{44.5}-10^{45.5}$ erg s$^{-1}$ and Eddington ratios of $0.07-0.47$.  
The spatially extended structure of the F200W stacked image may trace the stellar light from the host galaxies. The  \ha\ luminosity function indicates an increasing AGN fraction towards the higher \ha\ luminosities. 
We find possible evidence for clustering of BHAEs: two sources are at the same redshift with a \xj{projected} separation of 519 kpc; one BHAE appears as a composite system residing in an overdense region with three close companion  \ha\ emitters. Three BHAEs exhibit blueshifted absorption troughs indicative of the presence of high-column-density gas. \xlin{We find the broad-line and photometrically selected BHAE samples exhibit different distributions in the optical continuum slopes, which can be attributed to their different selection methods.} The ASPIRE broad-line \ha\ sample provides a good database for future studies of faint AGN populations at high redshift.

\end{abstract}


\section{Introduction}

Tremendous efforts have been made in the past decades to search for and characterize supermassive black holes (SMBHs) in distant bright quasars and active galactic nuclei  \citep[AGNs;][]{Fan2023}. Luminous quasars at $z>5$, 
and BH mass of $\sim 10^9M_\odot$, pose challenges to BH formation theories, with rapid mass assembly within a short amount of cosmic time \citep[e.g.,][]{Banados2017,Wang2021, Yang2021}.  Such a rapid growth with sustained episodes would erase the 
imprints of initial BH seeds \citep{Volonteri2010}, necessitating studies of high redshift low-mass BHs and low-luminosity AGNs. However, the knowledge of early low-mass BHs is very limited before the launch of James Webb Space Telescope (JWST).  
With the unprecedented infrared capabilities of JWST, we are now able to their early, low-mass, low-luminosity systems, closer to the early stage of BH growth and the predicted seed BH population \citep{Inayoshi2022}.  Faint AGNs at high redshift with smaller BH provide crucial constraints to the BH seeding, growth, and coevolution with galaxies \citep{Pacucci2023,Volonteri2023,Ding2023,Li2024}. 

In pursuit of the low-luminosity AGNs \citep[e.g.,][]{Onoue2023, Ubler2023, Furtak2023, Kocevski2023}, JWST reveals an abundant population of red compact sources at $z>4$ \citep[e.g.,][]{Barro2023, Labbe2023}, nicknamed as `little red dots' \citep[LRDs]{Matthee2023}.  \xj{Their number density is $10- 100\times$ higher than the extrapolation from quasars luminosity functions \citep[e.g.,][]{Kocevski2023}.} The nature of LRDs remains a mystery. JWST spectroscopy detects moderately luminous and broad \ha\ or $H\beta$ lines in the LRDs,  suggesting 
SMBHs with masses ranging from $\sim10^6$ to $\lesssim 10^9 M_\odot$ \citep{Kokorev2023, Kocevski2023, Furtak2023, Maiolino2023, Matthee2023, Greene2023, Harikane2023}. \xj{Most of them have SMBH masses} 1--2 dex lower than those of UV bright quasars at similar redshifts identified before the launch of JWST \citep{ Onoue2019, Yang2023}.  LRDs typically exhibit  V-shape spectral energy distribution \citep[SED,][]{Labbe2023, Furtak2023, Greene2023, Killi2023}, with the optical continuum redder than that in the UV in the $f_\lambda$ space (typically the UV continuum slope $\beta_{\rm UV} < 0$, the optical continuum slope $\beta_{\rm opt} > 1$).  Whether the UV emission of LRDs originates from host galaxies or the scattered light of AGNs is still under debate \citep{Labbe2023, Kocevski2023, Perez2024}. 

However, before a full understanding of the physical nature of LRDs, one of the main areas of uncertainty is the selection criteria.  Spectroscopic selections \citep{Kocevski2023, Matthee2023,Maiolino2023, Harikane2023} are based on the broad \ha\ or H$\beta$ emission lines, 
\xj{indicative of broad-line regions around BHs}.  On the other hand, photometric selections are usually based on LRDs' V-shape SED and their compactness and redness \citep{Labbe2023, Greene2023, Kokorev2024, Perez2024}. Some photometrically selected LRD 
candidates have been confirmed as brown dwarfs spectroscopically \citep{Greene2023, Hainline2023}. Without 
showing broadened Balmer emission lines or other AGN line diagnostics \citep{Scholtz}, 
we also cannot rule out the possibility of dusty starburst galaxies. \textit{JWST}/MIRI studies on photometrically selected LRDs show flattened spectral shape in the rest-frame near-infrared ($\gtrsim 1\mu$m), inconsistent with emission from AGN dust tori \citep{Williams2023,Perez2024}. \cite{Perez2024} claimed that these photometrically selected LRDs are more likely to be extremely intense dusty starburst galaxies with obscured AGNs contributing in the mid-infrared.  \xlin{It is unclear whether the photometrically selected LRDs and those broad-line selected ones through WFSS or NIRSpec share the same physical properties. Furthermore, it also deserves further exploration to confirm if these LRDs are indeed AGNs.
 }   


To better constrain the nature of low-luminosity AGNs, 
 including the so-called LRDs, and study BH mass assembly at high redshifts, \xj{we need a large sample of} broad-line selected faint AGNs.   In this work, we perform a systemic search for broad \ha\ line emitters in the ASPIRE program \citep{Wang2023, Yang2023}. ASPIRE is the JWST Cycle 1 grism program covering a large sky area, totaling $\approx 65$ hours over $\approx 275$ arcmin$^2$, \xj{over 25 differnt fields}.  Using \textit{JWST}/NIRCam WFSS \citep{Greene2016} in F356W ($R\approx 1600$), we can build up a flux-limited broad \ha\ emitter sample at $z=4-5$.  ASPIRE's large survey area \xj{over multiple fields} allows quantifying the number density and AGN fraction, effectively mitigating the cosmic variance.   To avoid confusion, we refer to our broad-line selected sample as broad \ha\ emitters throughout this paper (BHAEs for short),  rather than LRDs.

This paper is organized as follows.  We describe the JWST observations, data reduction, and sample selection in \S\ref{sec:data_and_sample}. In \S\ref{sec:result} we present the properties of our selected broad \ha\ emitter sample, including their spectral properties (\S\ref{sec:broadha}), BH masses and bolometric luminosities (\S\ref{sec:BH_Lbol}), UV and optical continuum slopes (\S\ref{sec:beta} and morphology (\S\ref{sec:morph}).  We analyze the \ha\ luminosity function in \S\ref{sec:LF}.  We discuss interesting individual sources in our sample in \S\ref{sec:discussion}.  Our results are summarized in \S\ref{sec:summary}.  Throughout this work, a flat $\Lambda$CDM cosmology is assumed, with $H_0 = 70 {\rm ~km~s^{-1}~Mpc^{-1} }$, $\Omega_{\Lambda,0} = 0.7$ and  $\Omega_{m,0}=0.3$. Distances are provided in the comoving frame if not specified. 

\section{Data and Sample}\label{sec:data_and_sample}


\subsection{JWST observations}\label{sec:JWSTdata}
The ASPIRE (A SPectroscopic survey of biased halos In the Reionization Era) program (GO 2078; PI F. Wang) targets 25 quasars at $6.5<z<6.8$ with \textit{JWST}/NIRCam imaging and wide field slitless spectroscopic (WFSS) data, with a total survey area of $\sim 11\times25$ arcmin$^2$. For each field, ASPIRE obtains 2834.5s exposures in NIRCam/F356W Grism R, accompanied with direct imaging of 1417.3s 
in NIRCam/F115W and F356W, and 2800s in  F200W. 
The data is reduced in the same way as described in \cite{Wang2023}. We summarize the main procedure below.  

The NIRCam images are reduced with the standard pipeline v1.10.2\footnote{\url{https://github.com/spacetelescope/jwst}}.    We utilize the reference files ({\tt jwst\_1080.pmap}) from version 11.16.21 of the standard Calibration Reference Data System\footnote{\url{https://jwst-crds.stsci.edu}} (CRDS)  for calibration.  The \textit{1/f} noise is subtracted on a row-by-row basis and column-by-column basis after {\tt CALWEBB} Stage 1.  After Stage 2, a master median background is generated using all available exposures from ASPIRE for each band. The master
backgrounds are then scaled and subtracted for individual exposures. All the images are mosaicked with a pixel scale of 0.031\arcsec\ and \texttt{pixfrac=0.8}. Their WCS are registered using the Gaia Data Release 2 catalog \citep{GAIA2016, Gaia2018}. \xj{we construct the photometric catalogs with SourceExtractor++\citep{SEpp}, using F356W as the detection images. We first measure the flux in elliptical apertures with a Kron factor of $k=1.2$ \citep{Kron1980}. Then we performed aperture correction for each source, multiplying the ratio of F356W flux within $k=2.5$ v.s. $k=1.2$ apertures. \xj{ SourceExtractor++ also provides the F356W half-light radii measurements.} More details about the photometric catalogs are presented in Champagne et al. in prep.} 

For NIRCam WFSS data, we subtract the \textit{1/f} noise only along columns. The median background models for WFSS are constructed using all available ASPIRE observations obtained at similar times and are scaled and subtracted from individual WFSS exposures.  The WCS of individual exposures is first aligned using the \texttt{CALWEBB} pipeline. We then measure the astrometric offsets between the direct images and the fully calibrated F356W-band mosaic, which are added to the spectral tracing model as described below. The spectral tracing models, dispersion models, and sensitivity functions are constructed as described in \cite{Sun2023}.  The spectral tracing model and the sensitivity functions used in this work are updated using the spectral traces of point sources observed in the LMC field (PID \#1076) and  Cycle-1 calibration programs (PIDs \#1536, \#1537, \#1538) respectively.  we construct and stack the 2D spectra for each source, using 2D spectra from individual exposures that are registered to a common wavelength and spatial grid following the algorithms in {\ttfamily{PypeIt}} \citep{Pypeit}. The pixel scales for the 2D spectra and 1D spectra described below are both resampled to be 10 \AA\ per pixel. 

We first extract 1D spectra using the optimal extraction algorithms and remove the background and continuum using median filters (see \citealt{Wang2023} for details). These continuum-removed 1D spectra are used for the initial emission line identification. After identifying the \ha\ emitter candidates, we re-extract the 1D spectra for our analysis, \xj{without continuum removal}. For those with contaminated 2D spectra, we mask regions of $\pm 5000~ {\rm km ~s^{-1}}$ around the \ha\ emission to avoid over-subtraction around the broad wing. We then run median filters iteratively along each row of the 2D spectra to construct continuum maps.  We re-extract 1D spectra from these continuum-removed 2D spectra using the optimal extraction algorithms.  For samples whose 2D grism spectra are not contaminated by nearby bright sources, we re-extract the 1D spectra based on their original grism data.  


\subsection{Sample selection}\label{sec:selection}

We systematically search for broad-line \ha\ emitters in the 25 ASPIRE fields over a total area of 275 arcmin$^2$.  We select compact red sources first with the following morphology and color cut:
\begin{itemize}
    \item[(1)]  F356W half-light radius 
    $<$ 0.2 arcsec;
    \item[(2)]  F200W $-$ F356W $>$ 0.75 mag.
\end{itemize}
The first requirement for compactness ensures that the source morphology does not broaden the \ha\ line profiles during the dispersion of grism spectra. The second criterion describes the F356W excess boosted by the strong \ha\ emission line. We define the color cut based on the broad-\ha\ sample in \cite{Matthee2023}. We then perform an emission line search for all the photometrically selected red compact sources and select objects with strong emission lines at a signal-to-noise ratio (S/N) \xj{of integrated line fluxes} greater than five. Careful visual inspection is performed to remove artifacts in the imaging and grism data, false detection by the search algorithms, and misidentified lines from contamination
of other sources along the dispersion direction. We then fit the line profiles of the emitter candidates with single and two-component Gaussian models. Broad \ha\ emitters are identified in two cases: (\romannum{1}) A single Gaussian model can well describe the line profile and it has an FWHM  $>$ 1000 km s$^{-1}$; or (\romannum{2}) the two-component model with an FWHM $>$ 1000 km s$^{-1}$ broad component performs better, resulting in $|\chi_{\rm red}^2-1|$ at least 0.5 smaller than that of single Gaussian models. The spectroscopic selection can be summarized as:
\begin{itemize}
    \item[(3)]  Emission lines with S/N \xj{of integrated line fluxes} $>$ 5;
    \item[(4)]  Robust broad components with FWHM$_{\rm broad}$ $>$ 1000 km s$^{-1}$.
\end{itemize}
To test our selection criteria, we generate a mock spectrum assuming a source with an effective radius of 0.2 arcsec and a single emission line with an intrinsic FWHM of 500 km s$^{-1}$. The mock emission line in the dispersed spectrum, \xj{broadened by both} the line spread function of NIRCam WFSS \xj{and the source morphology,} displays a Gaussian profile with FWHM of 715 km s$^{-1}$. \xj{This suggests that} with flux radii $<$ 0.2 arcsec and FWHMs of \ha\ broad components $>$ 1000 km s$^{-1}$, the \ha\ of our selected BHAE should have intrinsically broad lines.

\begin{figure*}[htbp]
    \centering
    \includegraphics[width=\columnwidth]{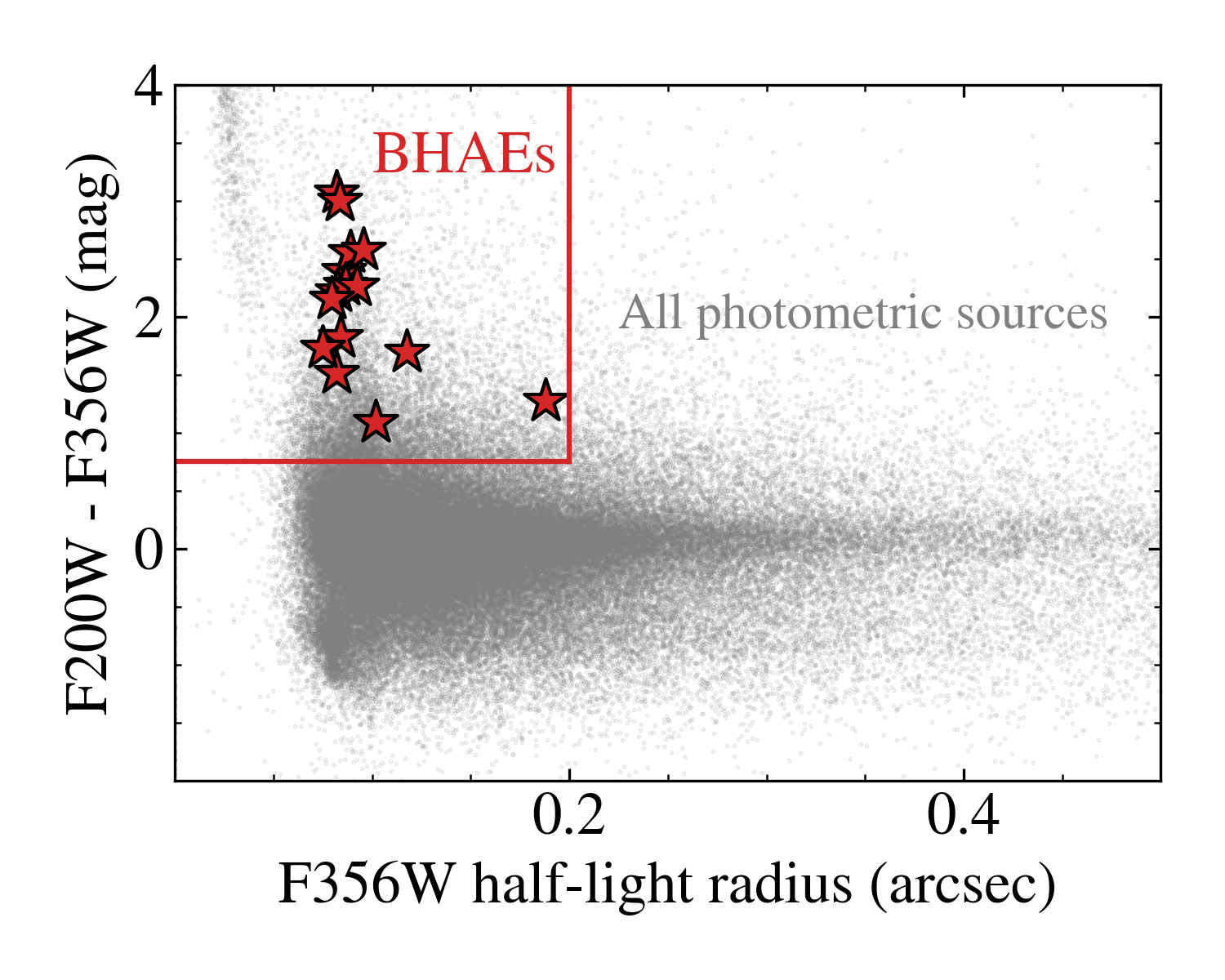}
     \includegraphics[width=\columnwidth]{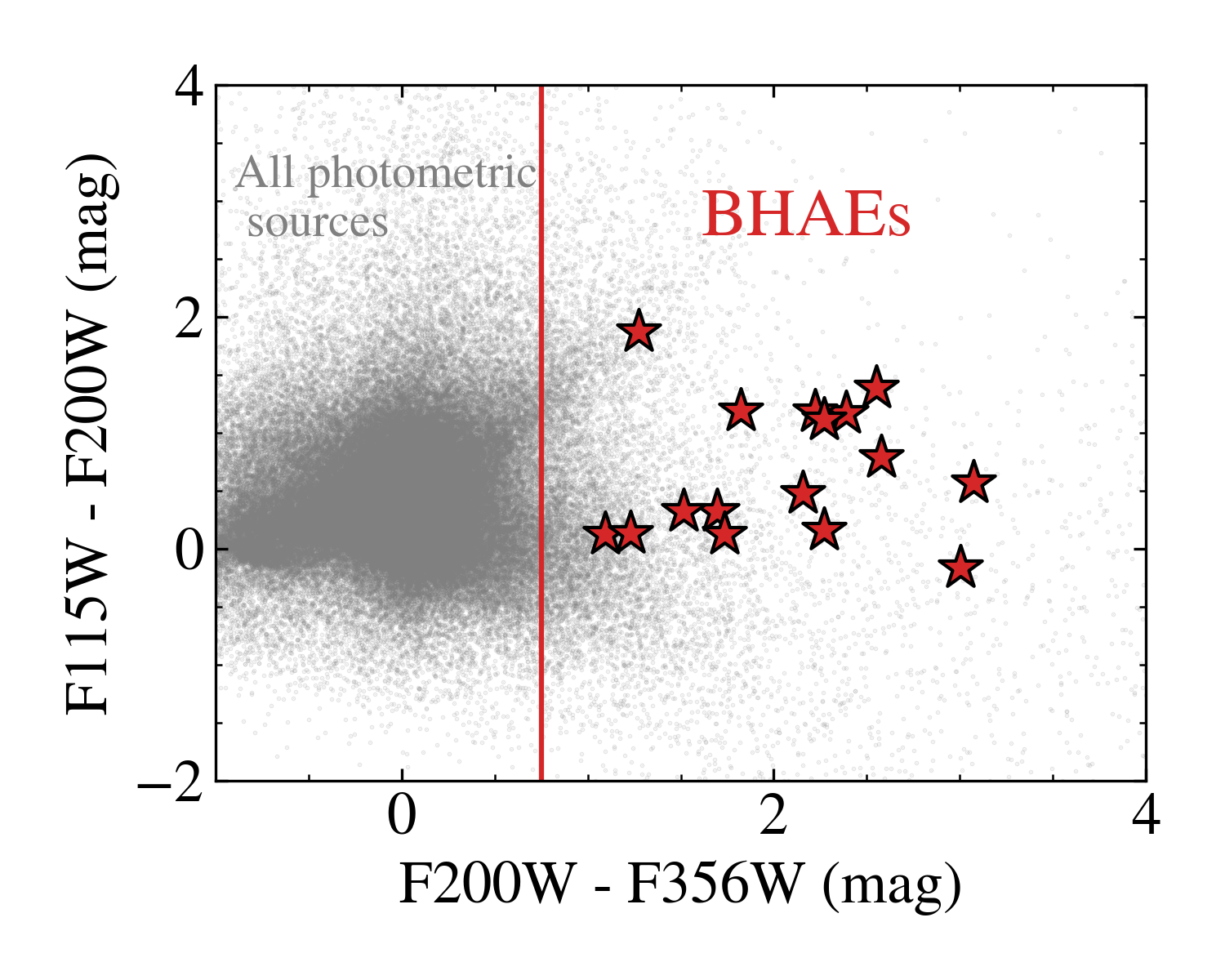}                
    \caption{The selection criteria of ASPIRE BHAE   (red stars) among all photometric sources in the 25 ASPIRE fields (gray dots). We label the applied color cut and size cut as red lines. 
    }
    \label{fig:selection}
\end{figure*}

We finally \xj{identify a total of} 16 BHAEs in the 25 ASPIRE fields. \xj{Figure \ref{fig:selection} presents our selection criteria and the 16 selected BHAEs. Their basic properties are listed in Table \ref{tab:basic}.} \xj{We note that our selection does not rely on the spectral energy distribution (SED) shape of sources \citep[e.g.,][]{Greene2023}. We specifically select compact, point-source-like BHAEs, rather than selecting all BHAEs from a complete \ha\ emitter (HAE) sample. The constraint of compactness maximizes the purity of our BHAE sample. 
}

\begin{table*} 
\centering
\begin{tabular}{ccccc}
    \hline
    ID & RA & DEC & $z$ & F356W mag \\
    \hline
    J0109M3047-BHAE-1 & $17.4772$ & $-30.8353$ & 4.3600$\pm$0.0006 & 24.647$\pm$0.008 \\
 
    J0218P0007-BHAE-1 & $34.6950$ & $0.0874$ & 4.2291$\pm$0.0004 & 24.459$\pm$0.011 \\
 
    J0224M4711-BHAE-1 & $36.0952$ & $-47.2007$ & 4.0639$\pm$0.0003 & 24.710$\pm$0.014 \\
  
    J0229M0808-BHAE-1 & $37.4083$ & $-8.1459$ & 4.3653$\pm$0.0002 & 24.804$\pm$0.018 \\
   
    J0229M0808-BHAE-2 & $37.3964$ & $-8.1954$ & 5.0369$\pm$0.0002 & 25.074$\pm$0.018 \\
     
    J0430M1445-BHAE-1 & $67.6898$ & $-14.7873$ & 4.0947$\pm$0.0003 & 23.398$\pm$0.004 \\
   
    J0910M0414-BHAE-1 & $137.7143$ & $-4.2213$ & 4.9102$\pm$0.0003 & 24.878$\pm$0.018 \\
 
    J0923P0402-BHAE-1 & $140.9658$ & $4.0545$ & 4.8688$\pm$0.0005 & 23.815$\pm$0.005 \\
 
    J1526M2050-BHAE-1 & $231.6589$ & $-20.8692$ & 4.1573$\pm$0.0005 & 23.860$\pm$0.006 \\
 
    J1526M2050-BHAE-2 & $231.6513$ & $-20.8566$ & 4.8708$\pm$0.0007 & 25.181$\pm$0.026 \\
 
    J1526M2050-BHAE-3 & $231.6710$ & $-20.8434$ & 4.8751$\pm$0.0005 & 25.477$\pm$0.018 \\
 
    J2002M3013-BHAE-1 & $300.6666$ & $-30.1814$ & 4.9461$\pm$0.0003 & 24.091$\pm$0.003 \\
   
    J2132P1217-BHAE-1 & $323.1515$ & $12.3108$ & 4.9250$\pm$0.0001 & 24.852$\pm$0.022 \\
    
    J2232P2930-BHAE-1 & $338.2275$ & $29.4940$ & 4.1349$\pm$0.0003 & 23.390$\pm$0.008 \\
    
    J2232P2930-BHAE-2 & $338.2381$ & $29.5225$ & 4.4693$\pm$0.0002 & 25.905$\pm$0.014 \\
   
    J2232P2930-BHAE-3 & $338.2325$ & $29.5094$ & 4.7061$\pm$0.0003 & 25.324$\pm$0.012 \\
    \hline
\end{tabular}
 \caption{The coordinate, redshift, and AB magnitude in F356W of the 16 LRDs in the ASPIRE fields. \texttt{BHAE} refers to \texttt{Broad H$\alpha$ emitters} throughout the paper. \xj{The redshift uncertainties are determined from the uncertainties in Gaussian fitting of \ha. The uncertainties in F356W magnitude are based on measurements. However, we apply an uncertainty floor of $\Delta m=0.05$ when calculating the continuum slope.}
 \label{tab:basic}
 }
\hspace{0.5cm}
\end{table*}

\begin{figure*}
    \centering
    \includegraphics[width=\columnwidth]{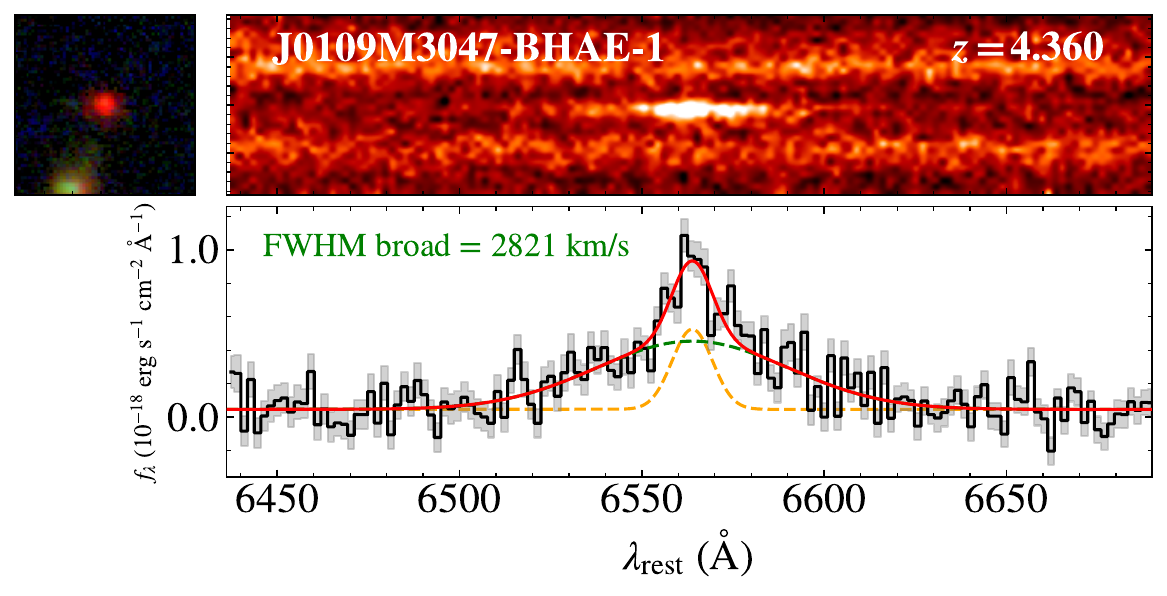}
     \includegraphics[width=\columnwidth]{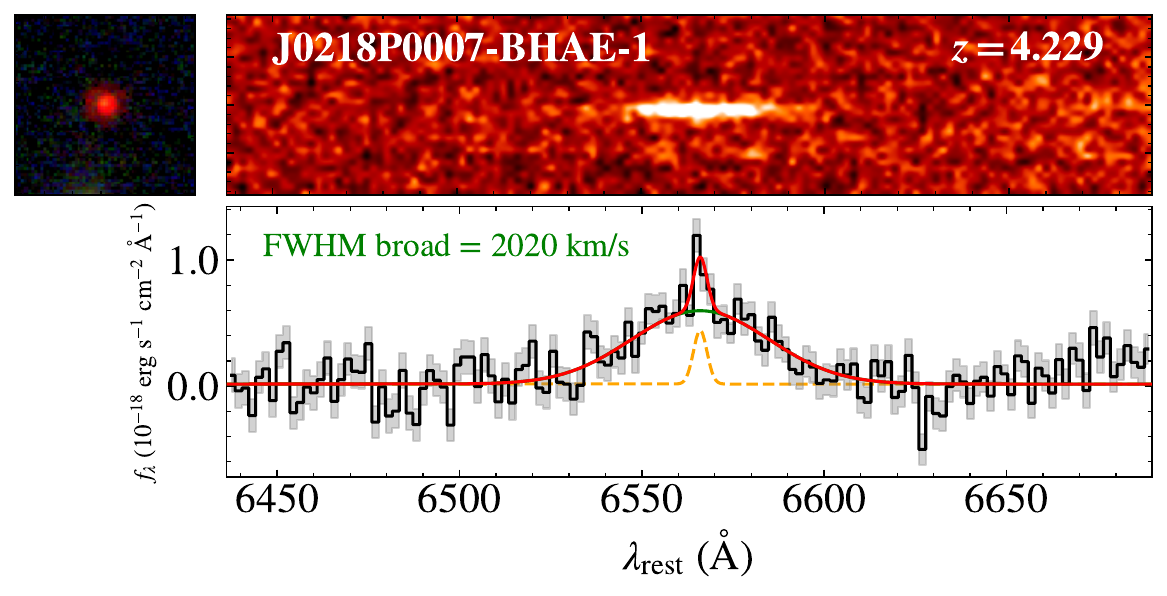}
     \includegraphics[width=\columnwidth]{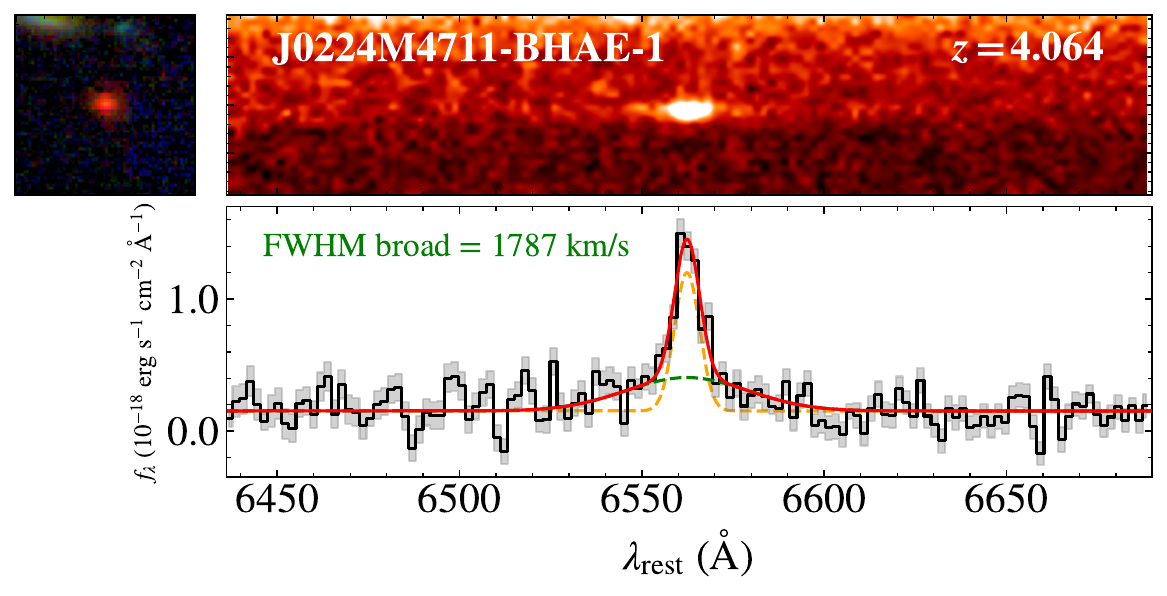}
     \includegraphics[width=\columnwidth]{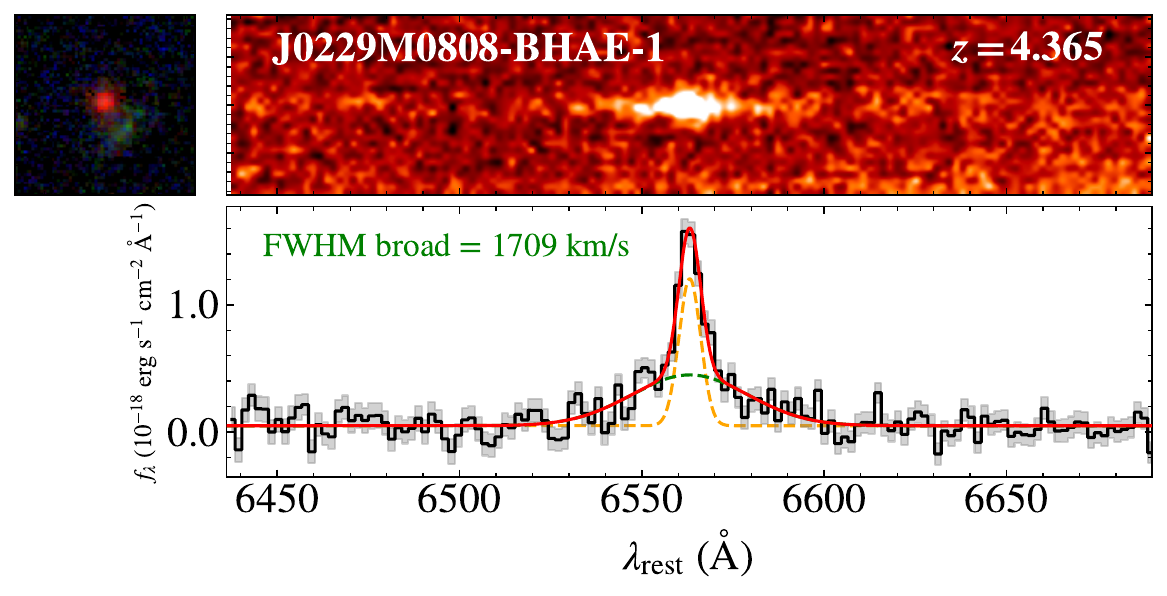}
     \includegraphics[width=\columnwidth]{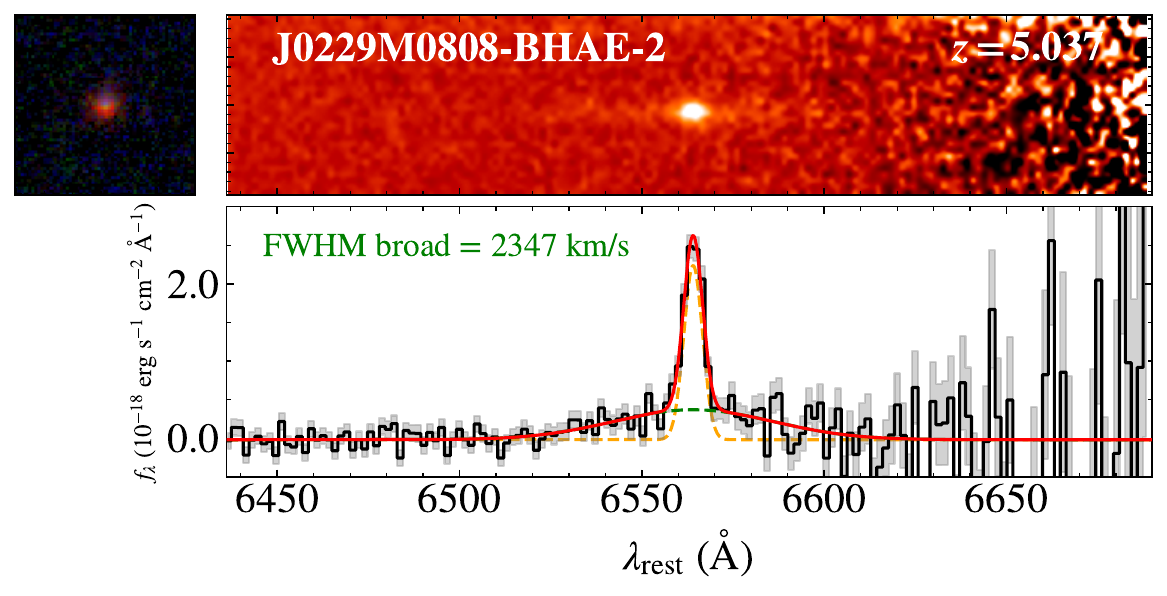}
     \includegraphics[width=\columnwidth]{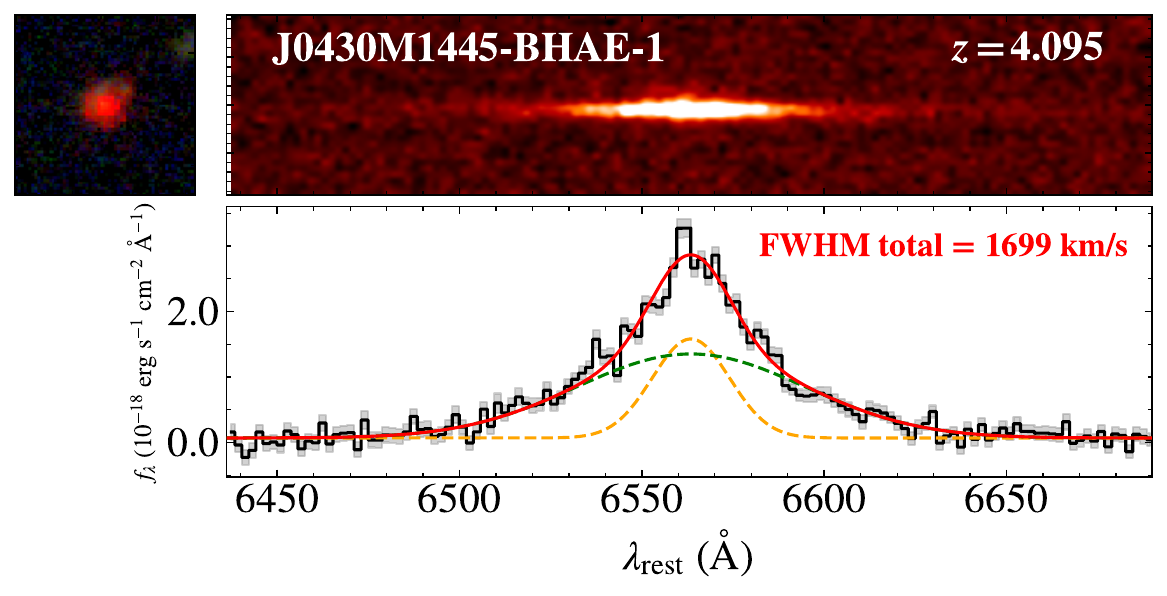}
     \includegraphics[width=\columnwidth]{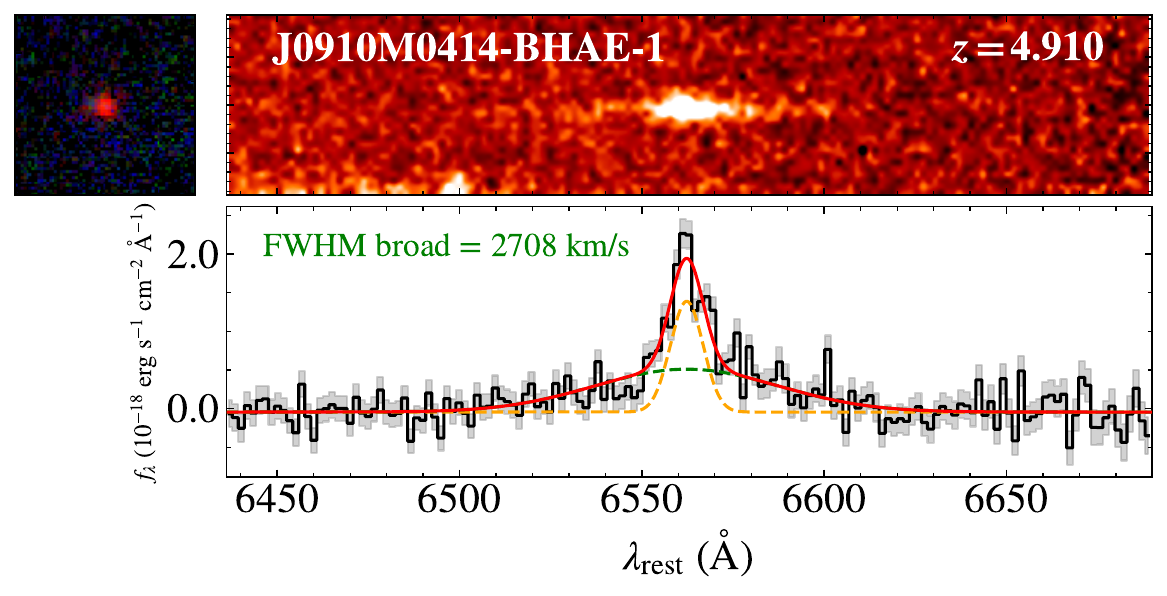}
     \includegraphics[width=\columnwidth]{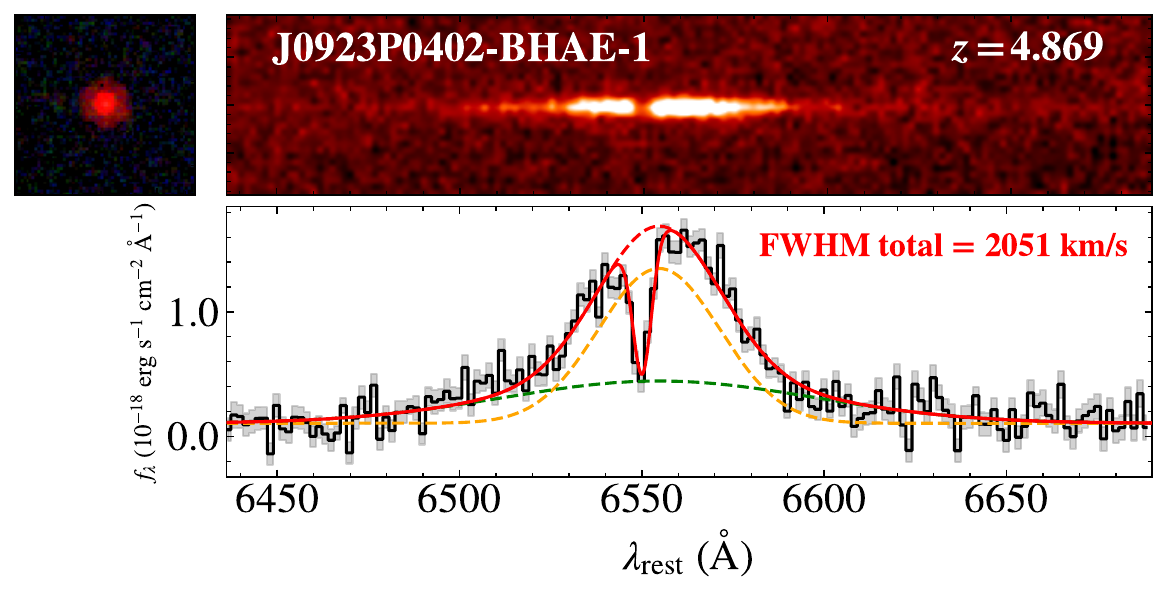}
     \includegraphics[width=\columnwidth]{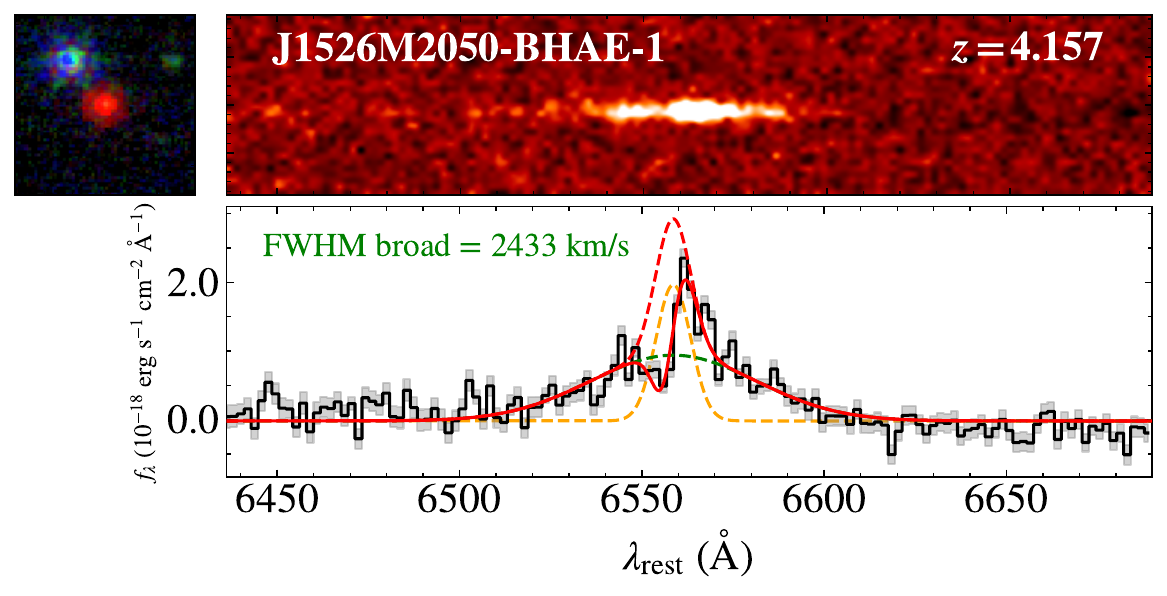}
     \includegraphics[width=\columnwidth]{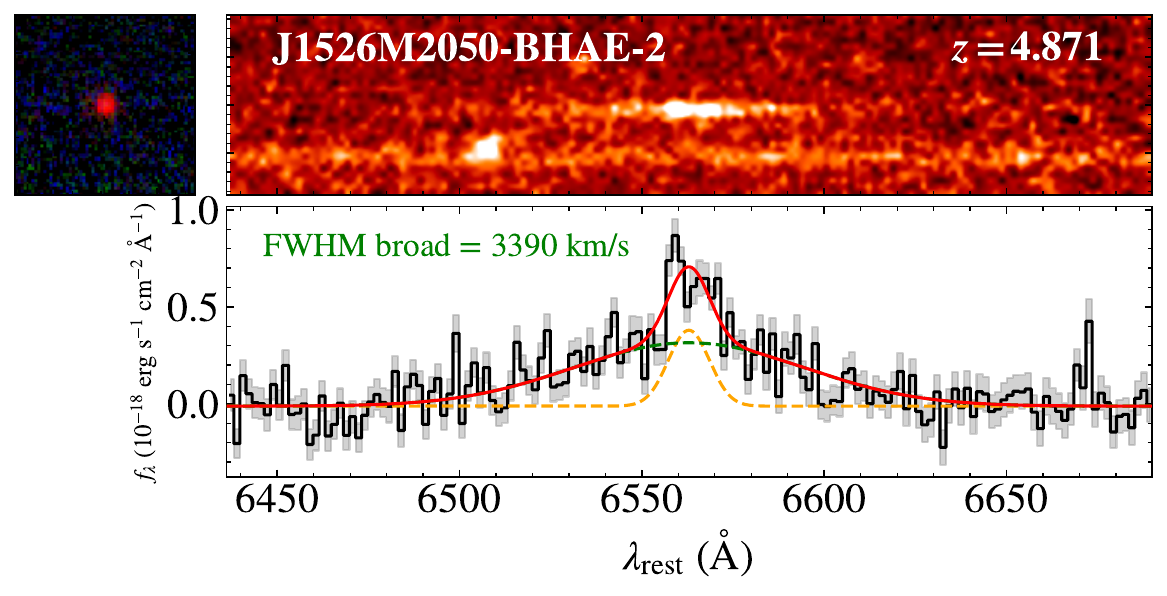}
    
    \caption{ The broad-\ha\ emitters identified in 25 ASPIRE fields. For each source, the upper left panel shows a 2\arcsec$\times$2\arcsec\ RGB thumbnail composed of \textit{JWST}/NIRCam F356W, F200W, and  F115W images.  The top panel shows the 2D grism spectrum with continuum and background removed. The bottom panel is the optimally extracted 1D spectrum (black) and the corresponding error spectrum (gray-shaded region). We show the total best-fit line profile as the red solid line, the narrow component as the orange dashed line, and the broad component as the green dashed line. For J0923P0402-BHAE-1, J1526M2050-BHAE-1, and J1526M2050-BHAE-3, we also show the \ha\ profiles without the absorption features as the red dashed lines.  \xj{J0923P0402-BHAE-1 and J0430M1445-BHAE-1 present two broad \ha\ components with FWHM$>$1000 km s$^{-1}$. We label the FWHMs of their total \ha\ emission, which are used to estimate the BH masses.}  }
    \label{fig:individual_Fig1}
\end{figure*}

\begin{figure*}
\centering
\gridline{\fig{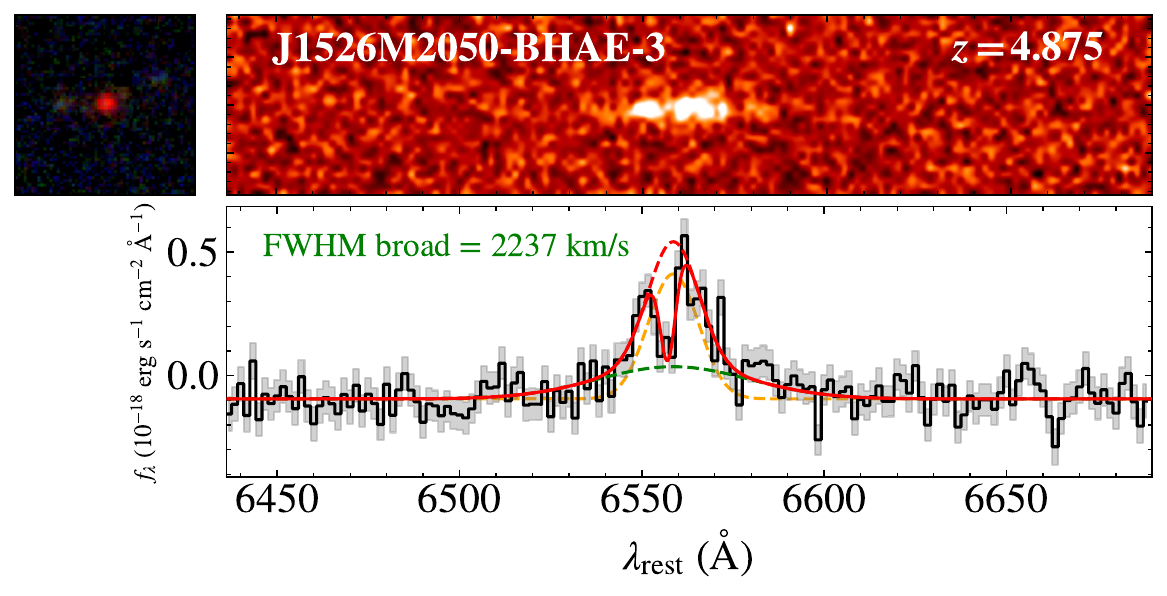}{\columnwidth}{}
          \fig{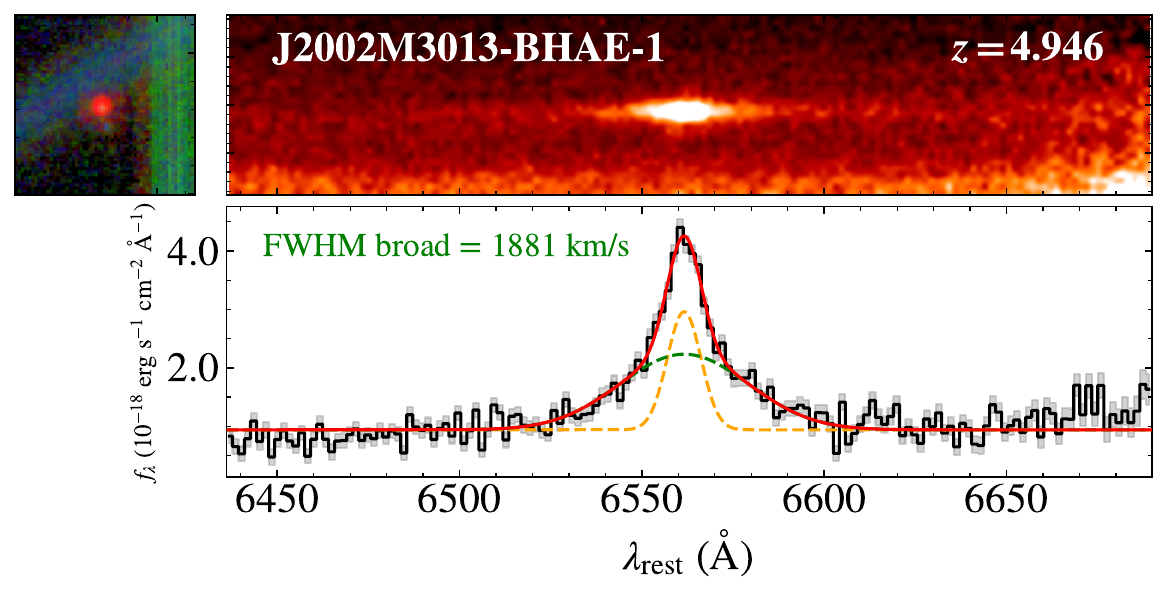}{\columnwidth}{}
          } 
\gridline{\fig{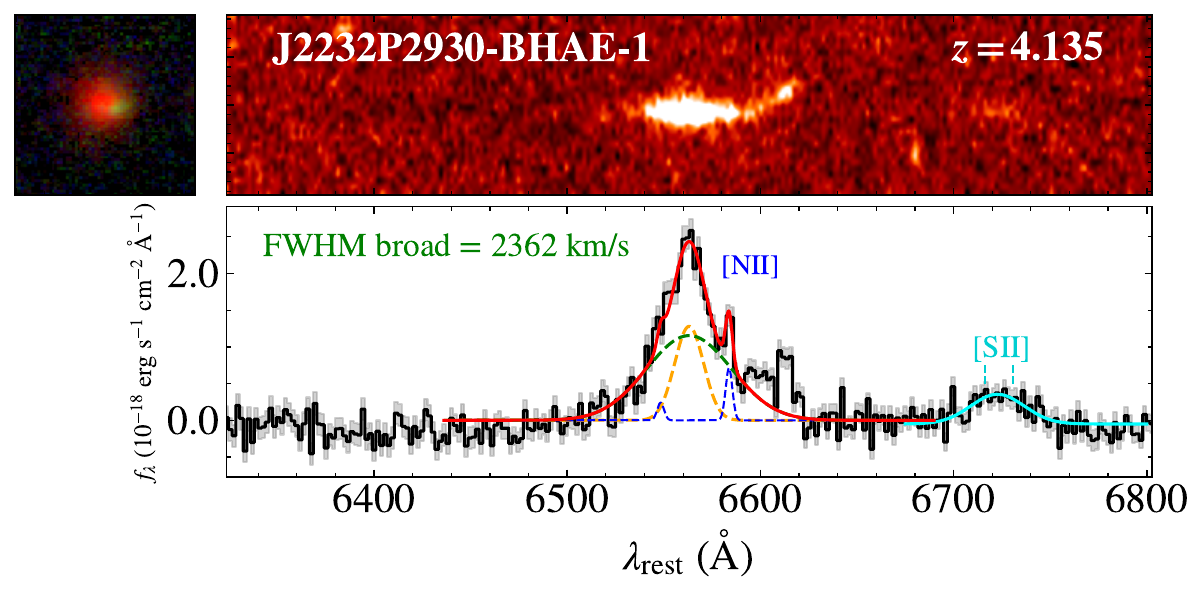}{\columnwidth}{}
          \fig{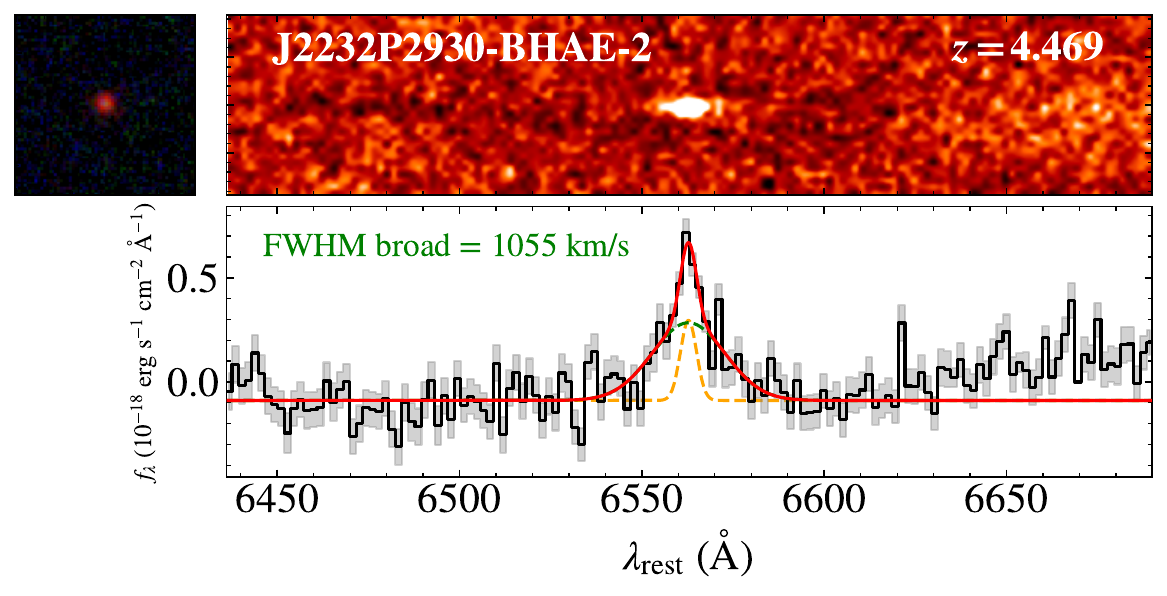}{\columnwidth}{}
          } 
\gridline{ 
        \fig{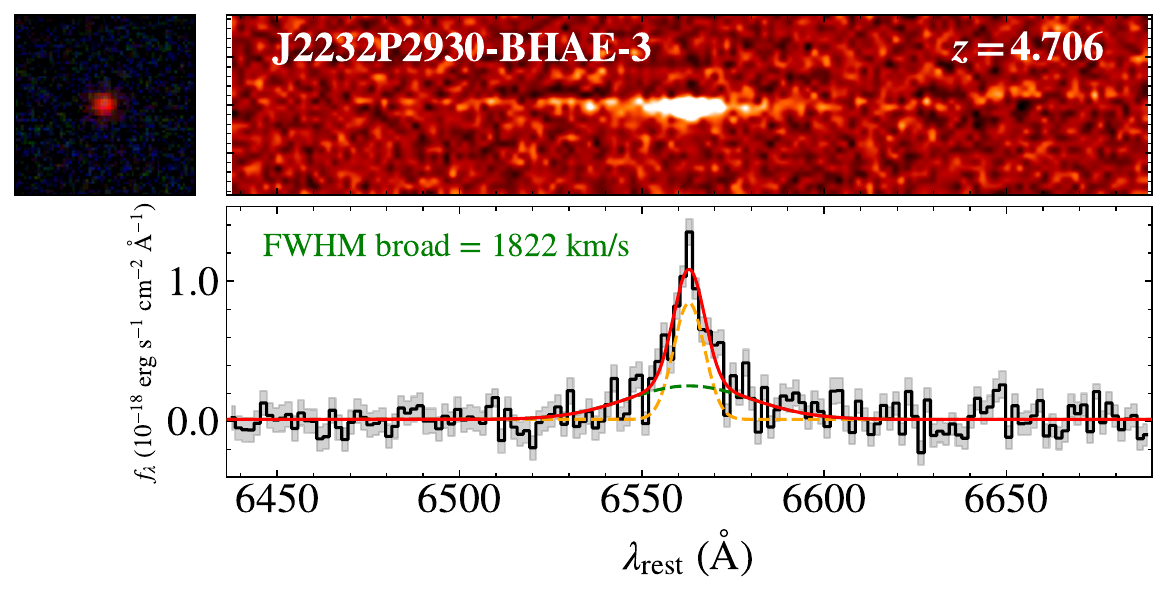}{\columnwidth}{} 
        \fig{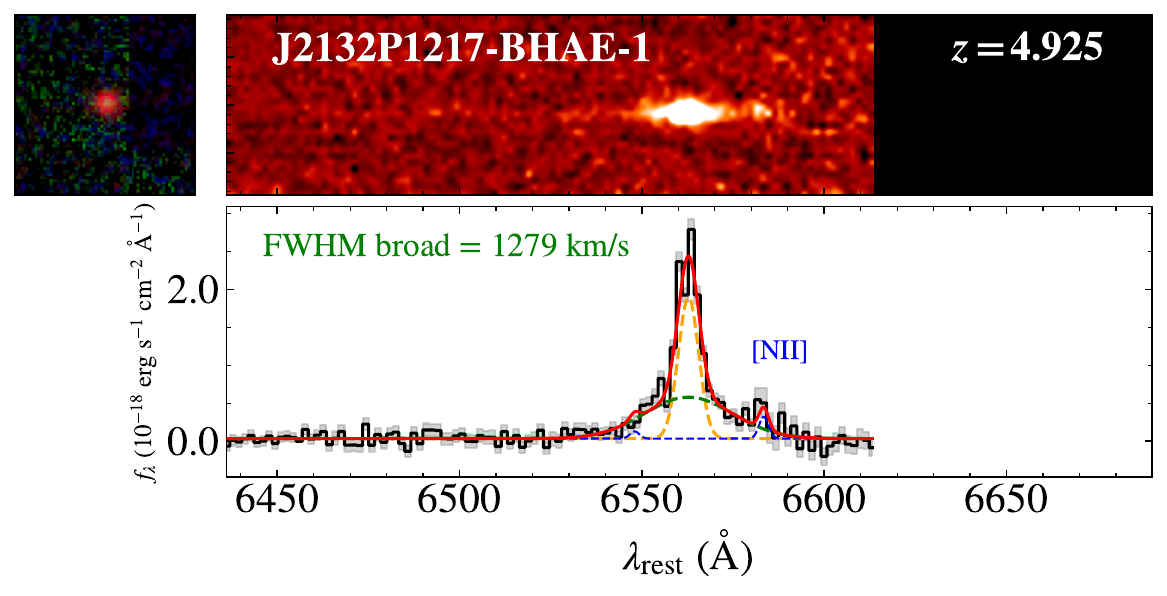}
        {\columnwidth}{} 
        }
\caption{Same as Figure \ref{fig:individual_Fig1}.  For J2232P2930-BHAE-1, we also show the [\ion{N}{2}]$\lambda\lambda6548,6583$ emission lines in blue and [\ion{S}{2}]$\lambda\lambda6718,6732$ in cyan. In the 2D spectrum of  J2232P2930-BHAE-1, the emission on its top right originates from a companion HAE at the same redshift (companion-2, see \S\ref{sec:J2232P2930-BHAE-1}).
\label{fig:individual_Fig2}
}
\end{figure*}

\section{Analyses and Results}\label{sec:result}
In this section, we present our analyses and results on a sample of 
16 ASPIRE BHAEs. 
We describe the properties of the broad \ha\ emission lines in \S\ref{sec:broadha} and derive the black hole masses and bolometric luminosities in \S\ref{sec:BH_Lbol}.  The UV and optical continuum slopes are discussed in \S\ref{sec:beta}. Finally, we study the morphology and \ha\ luminosity function of BHAEs in \S\ref{sec:morph} and \S\ref{sec:LF}.
\begin{table*}
\centering
\begin{tabular}{cccccc}
    \hline 
    ID & FWHM$_{\rm broad}$   & FWHM$_{\rm narrow}$ & $L_{\rm H\alpha, broad}$   & $L_{\rm H\alpha, total}$  &  $\rm EW_{\rm H\alpha, broad}$ \\
     & (km s$^{-1}$) & (km s$^{-1}$) & ($10^{42}$ erg s$^{-1}$) & ($10^{42}$ erg s$^{-1}$) & (\AA) \\
    \hline
    J0109M3047-BHAE-1 & 2821$\pm$351 & 559$\pm$119 & 5.08$\pm$0.55 & 6.33$\pm$1.81 & 630$\pm$119 \\
    
    J0218P0007-BHAE-1 & 2020$\pm$184 & --& 4.84$\pm$0.44 & 4.84$\pm$0.44 & 476$\pm$53 \\
    
    J0224M4711-BHAE-1 & 1787$\pm$472 & 306$\pm$50 & 1.70$\pm$0.37 & 3.13$\pm$0.87 & 227$\pm$53 \\
    
    J0229M0808-BHAE-1 & 1709$\pm$217 & 253$\pm$31 & 3.04$\pm$0.32 & 4.68$\pm$0.74 & 405$\pm$56 \\
    
    J0229M0808-BHAE-2 & 2347$\pm$388 & 202$\pm$19 & 5.65$\pm$0.88 & 9.50$\pm$1.64 & 818$\pm$189 \\
    
    J0430M1445-BHAE-1 & 1699$\pm$16 & --& 22.97$\pm$3.94 & 22.97$\pm$3.94 & 1206$\pm$125 \\
    
    J0910M0414-BHAE-1 & 2708$\pm$397 & 444$\pm$53 & 8.79$\pm$1.14 & 12.82$\pm$2.41 & 2333$\pm$1308 \\
    
    J0923P0402-BHAE-1 & 2051$\pm$132 & --& 21.63$\pm$6.87 & 21.63$\pm$6.87 & 1121$\pm$240 \\
    
    J1526M2050-BHAE-1 & 2433$\pm$260 & 471$\pm$63 & 9.20$\pm$0.80 & 13.21$\pm$2.95 & 656$\pm$105 \\
    
    J1526M2050-BHAE-2 & 3390$\pm$446 & 572$\pm$121 & 6.35$\pm$0.79 & 7.70$\pm$2.14 & 1493$\pm$655 \\
    
    J1526M2050-BHAE-3 & 2237$\pm$763 & 719$\pm$146 & 1.68$\pm$0.56 & 3.84$\pm$1.60 & 345$\pm$149 \\
    
    J2002M3013-BHAE-1 & 1881$\pm$166 & 459$\pm$52 & 14.52$\pm$1.11 & 20.47$\pm$3.58 & 1081$\pm$270 \\
    
    J2132P1217-BHAE-1 & 1279$\pm$173 & 235$\pm$23 & 4.14$\pm$0.43 & 7.43$\pm$1.12 & 480$\pm$80 \\
    
    J2232P2930-BHAE-1 & 2363$\pm$310 & 780$\pm$134 & 10.62$\pm$1.32 & 14.63$\pm$5.32 & 436$\pm$79 \\
    
    J2232P2930-BHAE-2 & 1055$\pm$149 & 136$\pm$51 & 1.87$\pm$0.22 & 2.30$\pm$0.80 & 733$\pm$182 \\
    
    J2232P2930-BHAE-3 & 1822$\pm$442 & 417$\pm$61 & 2.31$\pm$0.47 & 4.33$\pm$1.21 & 469$\pm$133 \\
    \hline
\end{tabular}
 \caption{The \ha\ properties of LRDs in the ASPIRE fields. FWHM$_{\rm broad}$ and FWHM$_{\rm narrow}$ is the full-width-half-maximum measured for the broad and narrow \ha\ components respectively. $L_{\rm H\alpha, broad}$ is the \ha\ luminosity of the broad component and  $L_{\rm H\alpha, total}$ is the \ha\ luminosity of the total \ha\ emission without absorption. $\rm EW_{\rm H\alpha, broad}$ is the rest-frame equivalent widths of the broad \ha\ emission. The narrow \ha\ component of J0218P0007-BHAE-1 is unresolved and thus we do not show its FWHM$_{\rm narrow}$. \xj{J0430M1445-BHAE-1 and J0923P0402-BHAE-1 exhibit two Gaussian components both with FWHMs$>$1000 km s$^{-1}$, so we measure the FWHM and luminosity of the entire \ha\ emission lines as the FWHM$_{\rm broad}$ and $L_{\rm broad}$.} 
 \label{tab: measurement}
 }
\end{table*}

\begin{table*}
 \centering
\begin{tabular}{cccccc}
    \hline
    ID & $M_{\rm UV}$ & $\beta_{\rm UV}$ & $\beta_{\rm opt}$ & log($M_{\rm BH}/M_\odot$) & log($L_{\rm bol}$/erg s$^{-1}$) \\
    \hline
    J0109M3047-BHAE-1 & $-17.21\pm0.31$ & $-0.07\pm0.74$ & $1.08\pm0.32$ & $7.86 \pm 0.11$ & $45.00^{+0.04}_{-0.04}$ \\
    
    J0218P0007-BHAE-1 & $-17.39\pm0.30$ & $0.05\pm0.69$ & $1.08\pm0.16$ & $7.55 \pm 0.08$ & $44.98^{+0.03}_{-0.04}$ \\
    
    J0224M4711-BHAE-1 & $-17.43\pm0.19$ & $-0.00\pm0.40$ & $0.54\pm0.17$ & $7.23 \pm 0.24$ & $44.59^{+0.08}_{-0.09}$ \\
    
    J0229M0808-BHAE-1 & $-19.15\pm0.10$ & $-1.47\pm0.23$ & $0.10\pm0.18$ & $7.31 \pm 0.12$ & $44.81^{+0.04}_{-0.04}$ \\
    
    J0229M0808-BHAE-2 & $-20.07\pm0.04$ & $-1.79\pm0.15$ & $-1.32\pm0.32$ & $7.72 \pm 0.15$ & $45.04^{+0.05}_{-0.06}$ \\
    
    J0430M1445-BHAE-1 & $-17.68\pm0.18$ & $0.35\pm0.38$ & $1.19\pm0.24$ & $7.72 \pm 0.04$ & $45.57^{+0.06}_{-0.07}$ \\
    
    J0910M0414-BHAE-1 & $-18.98\pm0.13$ & $-1.74\pm0.40$ & $-0.70\pm1.18$ & $7.94 \pm 0.13$ & $45.21^{+0.05}_{-0.05}$ \\
    
    J0923P0402-BHAE-1 & $-18.62\pm0.13$ & $-1.04\pm0.41$ & $1.73\pm0.61$ & $7.87 \pm 0.09$ & $45.55^{+0.11}_{-0.14}$ \\
    
    J1526M2050-BHAE-1 & $-18.04\pm0.15$ & $-0.13\pm0.35$ & $0.87\pm0.26$ & $7.85 \pm 0.10$ & $45.22^{+0.03}_{-0.03}$ \\
    
    J1526M2050-BHAE-2 & $-18.43\pm0.27$ & $-2.32\pm0.93$ & $1.19\pm0.96$ & $8.07 \pm 0.12$ & $45.09^{+0.04}_{-0.05}$ \\
    
    J1526M2050-BHAE-3 & $-17.15\pm0.27$ & $-0.75\pm0.79$ & $1.23\pm0.62$ & $7.43 \pm 0.31$ & $44.59^{+0.11}_{-0.15}$ \\
    
    J2002M3013-BHAE-1 & $-20.86\pm0.03$ & $-1.77\pm0.12$ & $-1.45\pm0.46$ & $7.71 \pm 0.08$ & $45.40^{+0.03}_{-0.03}$ \\
    
    J2132P1217-BHAE-1 & $-19.53\pm0.09$ & $-1.46\pm0.31$ & $-0.51\pm0.27$ & $7.11 \pm 0.12$ & $44.93^{+0.04}_{-0.04}$ \\
    
    J2232P2930-BHAE-1 & $-18.15\pm0.15$ & $1.17\pm0.32$ & $-0.47\pm0.25$ & $7.85 \pm 0.12$ & $45.28^{+0.04}_{-0.05}$ \\
    
    J2232P2930-BHAE-2 & $-18.37\pm0.10$ & $-1.79\pm0.27$ & $-0.06\pm0.43$ & $6.78 \pm 0.13$ & $44.63^{+0.04}_{-0.05}$ \\
    
    J2232P2930-BHAE-3 & $-18.08\pm0.17$ & $-1.20\pm0.47$ & $0.54\pm0.39$ & $7.31 \pm 0.22$ & $44.71^{+0.07}_{-0.08}$ \\
    \hline

\end{tabular}

 \caption{The black hole and galaxy properties of LRDs in the ASPIRE fields. $M_{\rm UV}$ is the absolute UV magnitude at the restframe 1500 \AA.  $\beta_{\rm UV}$ is the UV slope derived from  the F115W and F200W magnitudes. $\beta_{\rm opt}$ is the optical slope  derived from the F200W and F356W magnitudes, where we subtract the \ha\ emission flux in the F356W to obtain the continuum flux. The black hole mass $M_{\rm BH}$ and bolometric luminosity.  
 \xj{The black hole mass and bolometric luminosity values have been calculated from the observed \ha\ emission without dust correction, hence their intrinsic values might be higher.}
 \label{tab:derived}
 }
\end{table*}

\subsection{Broad \ha\ emission}\label{sec:broadha}

\begin{figure*}
    \centering
    \includegraphics[width=\textwidth]{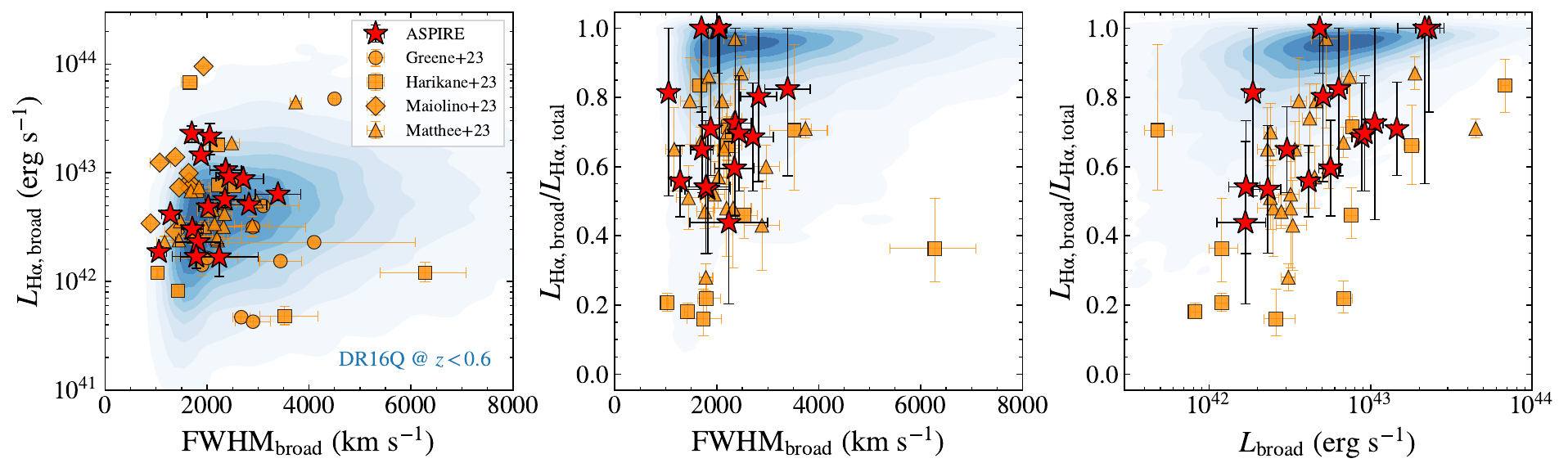}
    \caption{The properties of broad \ha\ emission lines of BHAEs. In the left panel, we show the relation between the  FWHM and the luminosity of the broad  \ha\  components. \xlin{We show the fraction of  \ha\ luminosity in the broad components as a function of the broad \ha\ FWHMs in the middle panel, and as a function of the broad \ha\ luminosity in the right panel.} The ASPIRE BHAEs are denoted as the red stars. We present the measurements of literature BHAEs  in \cite{Greene2023} (\xj{$4.5<z<8.5$}, orange circle), \cite{Harikane2023} (\xj{$4<z<7$}, orange square), \cite{Maiolino2023} (\xj{$4<z<7$}, orange diamond) and \cite{Matthee2023} (\xj{$4.2<z<5.5$}, orange triangle). The blue-shaded region denotes the distribution of $z<0.6$ quasars with \ha\ emission lines in SDSS spectra \citep{Wu2022}. 
    \label{fig:Lbroad}
    }
\end{figure*}

As described in \S\ref{sec:selection}, we fit the extracted 1D spectra with two-component Gaussian models for all 16 BHAEs. \xj{We do not impose any limits on the FWHM of the narrow-component Gaussian.} \xj{Figure \ref{fig:individual_Fig1} and \ref{fig:individual_Fig2} display the 16 BHAEs and their best-fit line profiles. The measurements are presented in Table \ref{tab: measurement}.} \xj{Among them, J0430M1445-BHAE-1 and J0923P0402-BHAE-1 have two broad components both with FWHM $>1000 {
\rm ~ km ~s^{-1}}$. The broad-line \ha\ emission is often not adequately modeled as a single Gaussian \citep{Shen2011}. We measure the FWHMs and luminosity of their entire \ha\ emission for BH mass estimates.} J0923P0402-BHAE-1, J1526M2050-BHAE-1, and J1526M2050-BHAE-3 show significantly blueshifted narrow absorption features of the \ha\ emission peaks \xj{(see the third to last and second to last panel of Figure \ref{fig:individual_Fig1}, and the first panel of \ref{fig:individual_Fig2})}. Similar features were also found in \cite{Matthee2023} and are interpreted as narrow \ha\ absorption.  Whether the features originate from \ha\ absorption or just complexes of multi-component emission lines remains uncertain. In this work, we assume they are \xj{caused} by \ha\ absorption and fit them as a single Gaussian. We adopt the total \ha\ luminosity values without absorption as the intrinsically emitted \ha\ luminosity for $L_{\rm H\alpha,  total}$. 
We note that the broad \ha\ fit is robust under the inclusion of the absorption, and exclude this component when measuring the broad \ha\ components.

Figure \ref{fig:Lbroad} presents the properties of broad \ha\ emission of ASPIRE BHAEs, together with literature high-redshift BHAEs  \citep{Greene2023,Matthee2023,Harikane2023,Maiolino2023} and low-redshift type-1 quasars observed by SDSS \citep{Wu2022}. We only present SDSS quasars at $z<0.6$ whose \ha\ lines are observable in SDSS spectra. As shown in the left panel of Figure \ref{fig:Lbroad}, the FWHMs and luminosities of the broad \ha\ emission of BHAEs are distributed similarly to those of type-1 quasars. It implies that BHAEs likely have broad-line regions similar to those of typical low-redshift type-1 quasars.  On the other hand, the fraction of \ha\ flux in the broad components, with a median of $0.70$, as shown in the \xlin{middle} panel of Figure \ref{fig:Lbroad}, is generally lower than that in most of 
low-redshift quasars (median 0.93). It indicates that the contributions from the narrow-line regions (NLRs) or host galaxies in these BHAEs to the total \ha\ luminosity (median $L_{\rm H\alpha, narrow}/L_{\rm H\alpha, total}=0.30$) are higher than those of low-redshift type-1 quasars (median $L_{\rm H\alpha, narrow}/L_{\rm H\alpha, total}=0.07$).  If the narrow \ha\ components largely originate from the host galaxies, the high values of $L_{\rm H\alpha, narrow}/L_{\rm H\alpha, total}$ suggest a higher star formation rate in the host galaxies of these BHAEs than those in typical low-redshift quasars.  \xlin{The right panel of Figure \ref{fig:Lbroad} shows that $L_{\rm H\alpha, broad} / L_{\rm H\alpha, total}$ generally increases with $L_{\rm H\alpha, broad}$.  It is likely a natural outcome of dust obscuration within BLRs, where BLRs of BHAEs with lower $L_{\rm H\alpha, broad}$  are subject to greater attenuation.}

\subsection{Black Hole mass and bolometric luminosity}\label{sec:BH_Lbol}

\begin{figure}
    \includegraphics[width=\columnwidth]{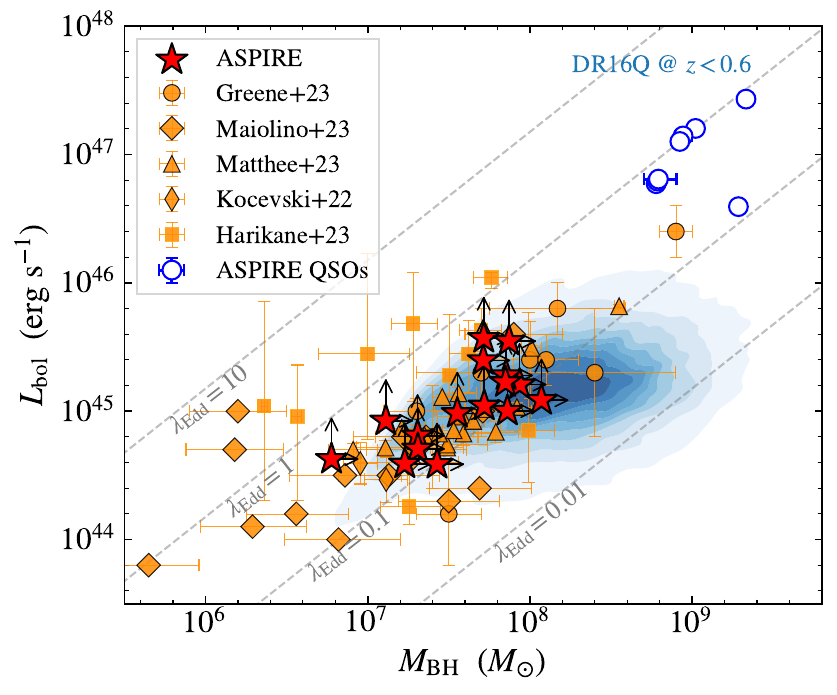}
    \caption{Black hole mass ($M_{\rm BH}$) versus bolometric luminosity ($L_{\rm bol}$). The red stars show the estimates in this work. The lengths of arrows indicate the changes of $M_{\rm BH}$ and $L_{\rm bol}$ when $A_V$ varies from 0 to 2. The orange circles, diamonds, triangles, and squares show the estimates from the literature BHAEs in \cite{Kocevski2023,Greene2023, Maiolino2023, Matthee2023, Harikane2023}. Note that the $L_{\rm bol}$ estimator used in  \cite{Harikane2023} is different as discussed in \S\ref{sec:BH_Lbol}. We present the distribution of low-redshift ($z<0.6$) quasars from \cite{Wu2022} as the blue-shaded region, with the depth demonstrating their number density. \xj{The blue-edged circles are measurements for $z>6.5$ quasars, deriving $M_{\rm BH}$ from $\rm H\beta$ emission lines and $L_{\rm bol}$ from $L_{\rm 5100}$.} We draw equal Eddington ratio lines (gray dashed) of $\lambda_{\rm Edd}=0.01, 0.1, 1, 10$.  
    \label{fig:Lbol_MBH}
    }    
\end{figure}

We estimate the bolometric luminosity (\Lbol) and black hole mass (\MBH) based on the broad \ha\ emission line as done in recent works \citep[e.g.,][]{Matthee2023,Harikane2023}.  Following \cite{Reines2013}, we estimate \MBH\ as:
\begin{equation}
\begin{array}{ll}
& \log _{10}\left(M_{\mathrm{BH}} / M_{\odot}\right)   = 6.57 +   \log _{10}(\epsilon) \\
& + 0.47 \log _{10}\left(L_{\mathrm{H} \alpha, \text { broad }} / 10^{42} ~\mathrm{erg} ~\mathrm{s}^{-1}\right)  \\
& + 2.06 \log _{10}\left({\rm FWHM}_{\rm broad} / 10^3 \mathrm{~km} \mathrm{~s}^{-1}\right).
\end{array}
\end{equation}
The geometric correction factor related to the broad line region, $\epsilon$, is assumed to be 1.075 \citep{Reines2015}. \xj{The BH estimator has an intrinsic uncertainty of 0.5 dex \citep{Reines2015}.} Note that, limited by our wavelength coverage, we derive the estimated \MBH\ values here without dust attenuation correction\xj{, as done in \cite{Matthee2023}}; the \MBH\ values could be even higher after correcting for the dust attenuation. 

We derive \Lbol\ by adopting the relation between the rest-frame 5100\AA\ luminosity ($L_{\rm 5100}$) and the AGN-induced \ha\ luminosity in \cite{Greene2005}, with the bolometric correction in \cite{Richards2006}:
\begin{equation}\label{eq:LLbol}
\begin{array}{ll}
L_{5100} &= 10^{44}  \left( L_{\rm H\alpha} / {{ 5.25\times 10^{42}} {\rm erg\ s^{-1}}} \right)^{1/1.157}  {\rm erg\ s^{-1}}\\
L_{\rm bol} &= 10.33 \times L_{5100}.
\end{array}
\end{equation}
where $L_{\rm H\alpha}$ in Equation \ref{eq:LLbol} should, include \ha\ from both AGN narrow line and broad line regions in principle. Since it is still not clear whether the narrow \ha\ components of BHAEs originate from AGN narrow line regions or the host galaxies, we use $L_{\rm H\alpha, broad}$ as the input.  The $L_{\rm bol}$ could be higher if the dust attenuation correction is applied or the contribution from AGN narrow-line regions is included. \xj{ 
We discuss the caveat of the \Lbol\ estimator in \S\ref{sec:ndensity}}. 
The Eddington ratio $\lambda_{\rm Edd}$ is defined as $L_{\rm bol}/L_{\rm Edd}$, where $L_{\rm Edd}$ \xj{is the theoretical maximum luminosity achievable when radiation pressure and gravity are balanced in a spherical geometry.}. We calculate $\lambda_{\rm Edd}$ following \cite{Trakhtenbrot2011}: 
\begin{equation}
    \lambda_{\rm Edd} = \frac{L_{\rm bol} / {\rm erg\ s^{-1}}}{ 1.5 \times 10^{38} M_{\rm BH} / M_\odot } .
\end{equation}
\xj{We list the derived physical properties of BHAEs in Table \ref{tab:derived}}. Figure \ref{fig:Lbol_MBH} shows the estimated \MBH\ and \Lbol\ of ASPIRE BHAEs  \xj{together with} 
literature BHAEs\footnote{\xj{We note that the $L_{\rm bol}$ estimator is different in \cite{Harikane2023}, the central values of $L_{\rm bol}$ are estimated using both the narrow and broad \ha\ components, the lower limits of $L_{\rm bol}$ are estimated using the broad \ha\ only, and the upper limits are estimated under the assumption of type-2 AGNs with [\ion{O}{3}] and H$\beta$. In this work, we use the broad \ha\ only, following \cite{Matthee2023}.  }}  and SDSS quasars. To display the effect of dust attenuation, we also show arrows associated with each BHAE. The lengths of the arrows indicate changes in \MBH\ and \Lbol\ with $A_V$ varying from 0 to 2.  The \MBH\ of ASPIRE BHAEs ranges from $10^{7}$ to $10^{8} M_\odot$, comparable to those of low-redshift quasars. The \ha\ converted \Lbol\ of BHAEs implies that the BHs are accreting at $\lambda_{\rm Edd} \approx 0.07-0.47$, with a median $\lambda_{\rm Edd}$ of 0.17.   \xj{The  $\lambda_{\rm Edd}$ of BHAEs are comparable to or slightly lower than those of $z>5$ quasars \citep[median values $\sim 0.3 - 0.8$, ][]{Shen2019, Farina2022, Mazzucchelli2023}. }

\subsection{UV and optical slopes}\label{sec:beta}

\begin{figure*}
    \centering
    \includegraphics[width=\textwidth]{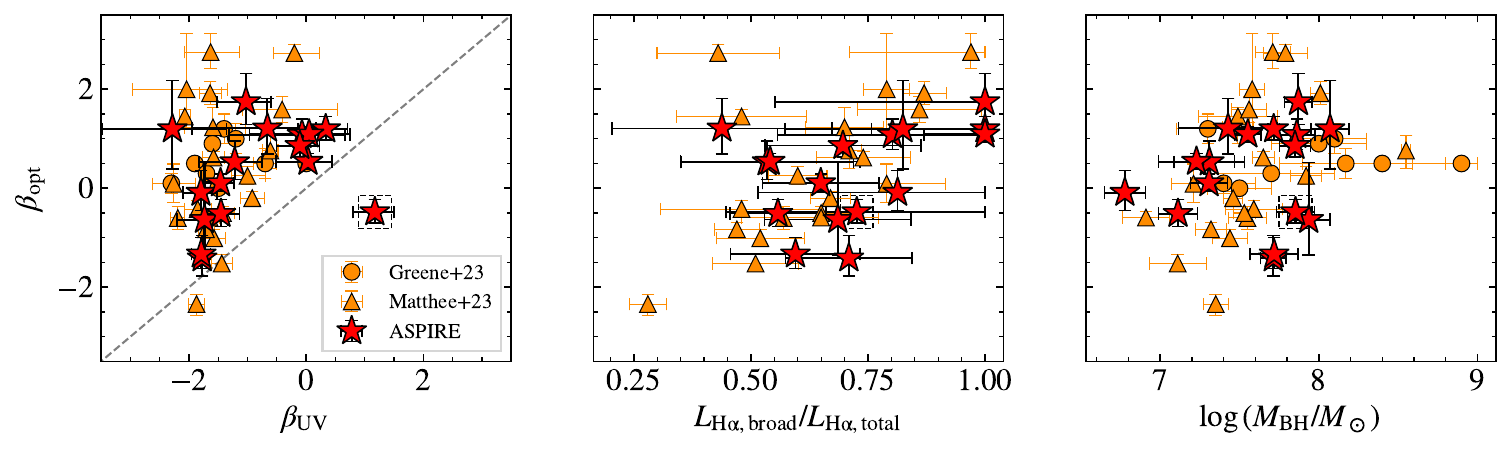}
    \caption{The relation between the optical slope ($\beta_{\rm opt}$) and UV slope ($\beta_{\rm UV}$), the fraction of \ha\ flux in the broad components ($L_{\rm H\alpha, broad}/L_{\rm H\alpha,total}$) and black hole mass ($M_{\rm BH})$. The measurements for ASPIRE BHAEs are denoted as red stars. The literature BHAEs in \cite{Greene2023} and \cite{Matthee2023} are shown as the orange circles and triangles. We label the source J2232P2930-BHAE-1 using a dashed square. More details about it are presented in \S\ref{sec:J2232P2930-BHAE-1}. }
    \label{fig:beta_opt}
\end{figure*}

The large difference in UV and optical slopes (\betauv, \betaopt, \xj{defined as $f_\lambda=\lambda^{\beta}$}) is one of the primary puzzles about the nature of so-called `little red dots'  \cite[e.g.,][]{Matthee2023, Killi2023}.  We approximate \betauv\ for all ASPIRE BHAEs by fitting the F115W and F200W photometry with a power law, which spans over rest-frame 2000~\AA\ -- 3600~\AA\ at $z\sim 4.5$. \xj{We assume that there are no strong emission lines in the two filters.} The  values of \betaopt\ are estimated using F200W and emission-line subtracted F356W photometry.  
\xj{We estimated the uncertainties of \betauv\ and \betaopt\ through  Monte Carlo (MC) method. This involved generating mock photometry based on the measurements and uncertainties and then recalculating the beta values $10^4$ times. An uncertainty floor of 0.05 mag is applied. The derived slopes are listed in Table \ref{tab:derived}}.

Figure \ref{fig:beta_opt} displays the relation between \betaopt\ and \betauv, the fraction of \ha\ flux in the broad component  $L_{\rm H\alpha, broad} / L_{\rm H\alpha,total}$  and \MBH.  {Most of the BHAEs  show redder \betaopt\ than \betauv\ except J2232P2930-BHAE-1.  J2232P2930-BHAE-1 is the most extended source in our sample with bright and diffuse \ha\ emission (see \S\ref{sec:J2232P2930-BHAE-1} for more details).}  We do not see a correlation between \betaopt\ and $L_{\rm H\alpha, broad} / L_{\rm H\alpha,total}$ in ASPIRE BHAEs.  The Kendall's $\tau$ correlation analysis on ASPIRE BHAEs yields a correlation coefficient of $r=0.27\pm0.22$ and $p$-value of $0.2_{-0.2}^{+0.4}$. Likewise, no strong correlation can be concluded from the \betaopt-\MBH\ relation in ASPIRE BHAEs. The Kendall's $\tau$ correlation analysis between \betaopt\ and $\log M_{\rm BH}$ of ASPIRE BHAEs shows $r=0.19_{-0.17}^{+0.18}$ and $p$-value of $0.3_{-0.3}^{+0.4}$.  {Our results are not consistent with the tight correlation seen in \cite{Matthee2023}, who interprets the correlations as transitions from star formation-dominated phase with small \MBH\ to dusty AGN-dominated phase with massive BHs. Our results suggest that the evolutionary track of these high-redshift AGNs might be non-linear or not identical among individual objects. \xj{However, we caution that the \betauv\ and \betaopt\ values in both this work and \cite{Matthee2023} are calculated based on only two broad bands. The UV and optical continuum shapes may not be simple power laws, so the estimates with two broad bands can be highly uncertain. Future \textit{JWST}/NIRSpec Prism \xlin{and Grating} spectra of larger BHAE samples are required to test the evolutionary scenario depicted in \cite{Matthee2023}. }  

\subsection{Morphology}\label{sec:morph}
Limited by the depth of ASPIRE observations, it is difficult to identify the host or companion galaxies of individual BHAEs.  To detect possible extended emission, we stack the images of 14 BHAEs in the ASPIRE sample. Among the total 16 ASPIRE BHAEs, J2232P2930-BHAE-1 is not included, because it is a composite system presenting significant extended \ha\ emission (\S\ref{sec:J2232P2930-BHAE-1}). 
J2002M3013-BHAE-1 is not included, because its image is severely contaminated by bright spikes of saturated stars nearby.  We generate 3\arcsec$\times$3\arcsec\ cutouts of the three bands for each broad \ha\ emitter and normalize the images to the value of the central brightest pixels. We stack the normalized cutouts \xj{and obtain the 3$\sigma$ clipped mean}. We use the median value of the outskirt 10 pixels in the stack of each band as the background level and subtract it from the final stack.  For cross-checking, we also perform PSF subtraction for each AGN using \texttt{Galfit} \citep{Peng2002}. The PSF models in each band are built \xj{by stacking} \textit{gaia} stars in the 23 ASPIRE fields (see more details in Champagne et al. in prep).  We subtract the PSF models within the central 8-pixel $\times$ 8-pixel regions of the cutouts. The PSF-subtracted images are stacked in the same way, normalized using the central brightest pixels of the original cutouts. We calculate the radial profiles of the two stack versions. The calculation of PSF-subtracted radial profiles starts at $r=1\ {\rm pixel}\ (0.031\ \rm arcsec)$, to avoid the 2-pixel$\times$2-pixel over-subtracted region in the centers. To compare the radial profiles of stacked images with the PSF profiles, we normalize all the profiles to their maximum values, so that the maximum values are unity at the minimum radius. To quantify the deviation of stacked images from the PSF shape, we define the deviation as
\begin{equation}\label{eq:deviation}
    \sigma(r) = \frac{S(r) - S_{\rm PSF}(r)}{\delta S(r)}
\end{equation}
where $\sigma(r)$ is the extent of the $S(r)$ profile deviation from the PSF profile $S_{\rm PSF}(r)$ with respect to the uncertainty $\delta S(r)$. For instance, $\sigma=1$ means that the radial profile $S(r)$ deviates from the PSF shape at a level of one sigma.

Figure \ref{fig:stack} shows the stacked images of BHAEs and the corresponding radial profiles.   
With the longest exposure time in F200W, both the F200W stack of original images and PSF-subtracted images present extended morphology. The radial profiles of the two F200W stacked images both deviate significantly from the shape of PSFs, with $\sigma\approx3.5$ at 0.1 arcsec and  $\sigma\approx1$ at $0.4$ arcsec. The stacked F200W extends to $\sim 2.6$ physical kpc (pkpc) at $z=4.5$.  The F115W and F356W stacks are also tentatively extended with $\sigma\gtrsim1$ within $r<0.3-0.4$ arcsec.   {The presence of extended structure in F200W, corresponding to $\sim$ restframe 3600 \AA\ at $z\sim4.5$, \xj{ is likely caused by the stellar emission} from the host galaxies. 

\begin{figure*}[htbp]
    \centering
    \includegraphics[width=\columnwidth]{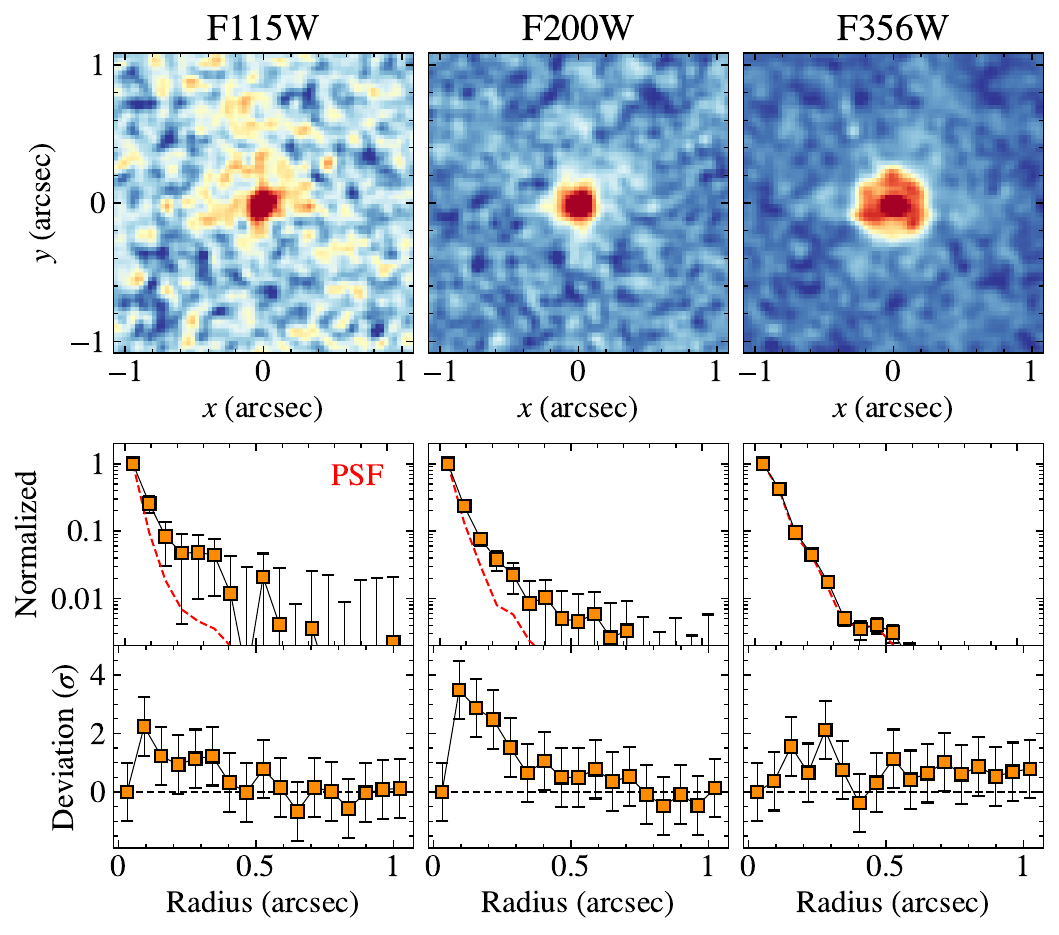}
    \includegraphics[width=\columnwidth]{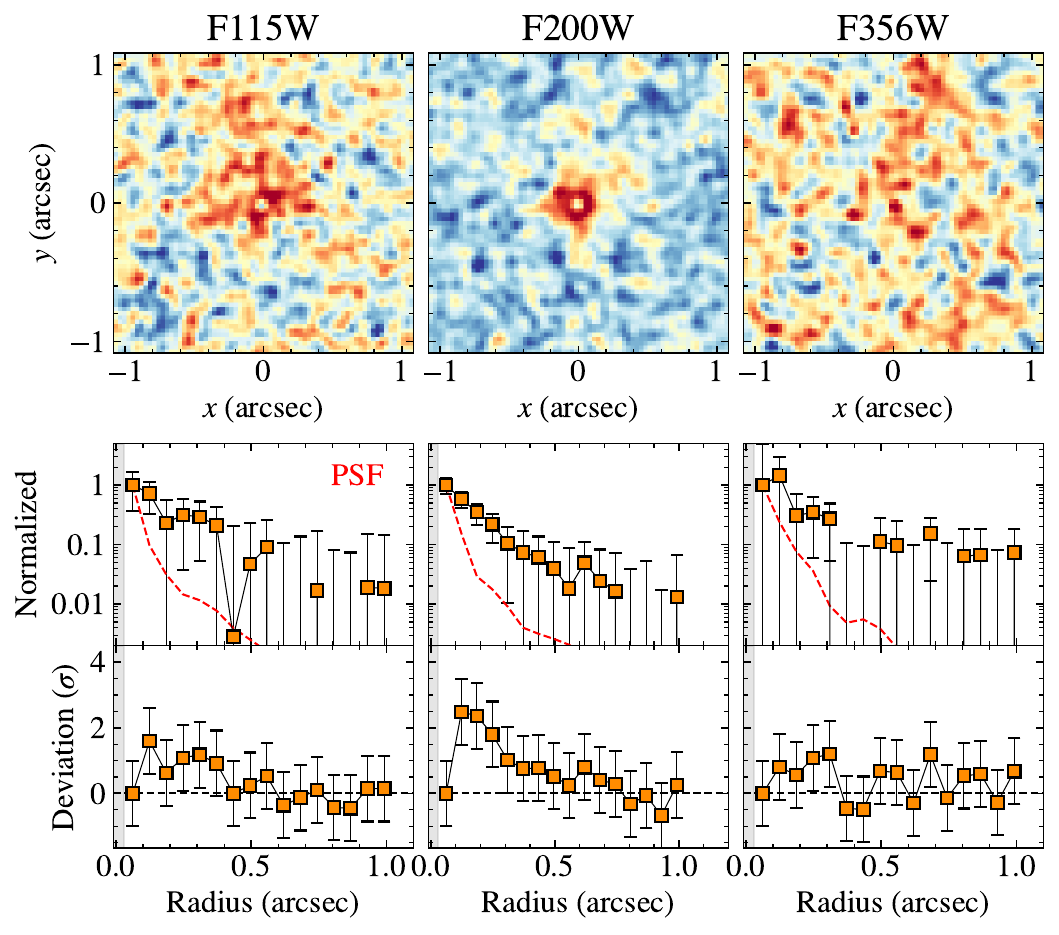}
    \caption{\textit{Left:} stack of images of 13 BHAEs  in NIRCam F115W, F200W and F356W, normalized to the value of central pixels. The top panel shows the 3\arcsec$\times$3\arcsec\ stack of each band. In the middle panel, the radial profiles of the AGN stacks are shown as orange squares, and the radial profiles of PSFs are shown as red dashed lines. The bottom panel shows the deviations of the stacked AGN radial profiles from the corresponding PSFs as defined in Equation \ref{eq:deviation}. \textit{Right:} the stack of PSF-subtracted images of 13 BHAEs, also normalized by the value of central pixels in the original images. The radial profiles of the AGN stacks and the deviation from the PSFs are shown in the middle and bottom panels. The calculation of PSF-subtracted radial profiles starts at $r=1\ {\rm pixel}\ (0.031\ \rm arcsec)$, to avoid the 2-pixel$\times$2-pixel over-subtracted region in the centers. The $r<1$ pixel ranges are grey-shaded in the middle and bottom panels. }
    \label{fig:stack}
\end{figure*}

\subsection{Luminosity function}\label{sec:LF}

The large survey area of ASPIRE program on $275$ arcmin$^2$ (25 ASPIRE fields)
allows us to measure the number density of broad \ha\ line-selected BHAEs at the bright end and reduce the effect of cosmic variance.  We do not calculate the UV luminosity function (LF) of BHAEs since it is still uncertain whether the UV components of BHAEs originate from the host galaxies or the scattered light from AGNs. Instead, we focus on the \ha\ LF using the total \ha\ flux (narrow + broad) directly measured from the observations. 

To calculate the total survey volume, we first compute the effective survey area for each BHAE, where the noise levels enable us to reliably identify the broad wing of the \ha\ emission lines. To do this, we need to ascertain the maximum root mean square (RMS) value for the identification of the broad \ha\ lines. We generate a series of mock 2D grism spectra. We convolve the two-component Gaussian model of each ASPIRE AGN with the line spread function of NIRCam WFSS along the wavelength axis and the F356W PSF along the spatial axis. We then add noise to the 2D mock spectrum and extract the 1D spectrum using the boxcar algorithm.  We fit the extracted 1D spectrum using a single Gaussian model and a two-component Gaussian model. If the two-component Gaussian model yields a S/N$>3$ and $|\chi^2_{\rm red}-1|$ value that is 0.1 smaller than that of the single Gaussian model, the broad \ha\ line is successfully identified. We progressively increase the noise levels added to the mock 2D spectra until the broad \ha\ components cannot be characterized. The noise values at this stage are the maximum RMS for the broad \ha\ identification. The survey volume is computed following the procedure in \cite{Sun2023}.  We define a series of redshifts spanning from 3.6 to 5.1, with \ha\ emission lines residing at 3--4 $\mu$m. We compute the effective sky area based on the spectral tracing and grism dispersion models for a specific redshift and construct the RMS maps using continuum-removed WFSS  \texttt{cal} files.  The maximum sky area at this redshift is determined as the area on the RMS map with RMS noise smaller than the maximum RMS for \ha\ broad line identification.  The maximum survey volume ($V_{\rm max}$) for each BHAE is integrated from the maximum sky area
across  $z=3.6-5.1$ over all the 25 ASPIRE fields. The median $V_{\rm max}$ for our AGN sample is 911561 cMpc$^3$. 

The \ha\ LF of BHAEs  is then computed using the direct 1/$V_{\rm max}$ method \citep{Schmidt1968}: 
\begin{equation}\label{eq:HaLF}
\Phi(L)=\frac{1}{d \log L} \sum_i \frac{1}{C_i V_{\max , i}},
\end{equation}
\xj{where $L$ is the \ha\ luminosity of each LF bin,  $C_i$ accounts for the completeness correction, and $V_{{\rm max},i}$ is the maximum survey volume for the $i-$th source.}
\xlin{To measure completeness, we randomly draw noise values from the effective RMS map as described above. We generate 2D mock grism spectra for each BHAE based on the two Gaussian models, adding the selected RMS values. We then re-extract the 1D spectrum following the procedure in Section \ref{sec:data_and_sample}. We run 1000 mock realizations for each BHAE, re-fitting the two-component Gaussian model to the extracted spectra. The completeness is defined as the fraction of the realizations in which the broad \ha\ components are successfully characterized according to the criteria mentioned above.  Additionally, we perform 1000 MC experiments for each luminosity bin to calculate the uncertainties, propagating the uncertainties of line fluxes to the LF. In each MC run, we account for the Poisson noise associated with small number statistics following the prescription in \cite{Gehrels1986}.  We also take into account the potential incompleteness caused by the color cut (see Section \ref{sec:selection}). We generate $10^4$ mock BHAE optical spectra assuming uniformly distributed $L_{\rm broad}$ in log space with $\log(L_{\rm broad}/ \rm erg ~ s^{-1}$ from 41.5 to 44, $L_{\rm broad} / L_{\rm tot}$ from 0.1 to 1, \betaopt\ frin -2.5 to 2.5,  $\rm FWHM_{\rm broad}$ from 1000 to 3000 km s$^{-1}$, $\rm FWHM_{\rm narrow}$ from 500 to 900 km s$^{-1}$, F200W magnitude from 25.5 to 27 mag, and redshift within $z=4-5$. We measure the F200W-F356W color excess for these mock BHAEs. We find that, for the LF of the total \ha\ emission, the color cut only excludes 8.6\% of BHAEs in the faintest \ha\ luminosity bin ($L_{\rm tot} \approx 10^{42.5} \rm erg ~ s^{-1}$), and is nearly complete in more luminous bins. For the broad \ha\ LF, the fraction is 12.6\%  in the faintest bin ($L_{\rm broad} \approx 10^{42.3} \rm erg ~ s^{-1}$) and also complete in more luminous bins.  This minor incompleteness is insignificant when considering the substantial uncertainty and the completeness correction due to the background RMS. We do not correct the incompleteness caused by the color cut because the intrinsic distribution of the physical parameters remains unclear. } 

Figure \ref{fig:HaLF} shows the \ha\ LF for ASPIRE BHAEs and the number density of each \ha\ luminosity bin is listed in Table \ref{tab:HaLF}. The number density of BHAEs are much lower than star-forming galaxies at similar redshifts.  The fraction of BHAEs at each luminosity bin is computed as the number density of BHAEs over the number density of Lyman break galaxies as derived in \cite{Bollo2023}.  The fractions increase strongly with $L_{\rm H\alpha}$, from $2\%$ at $\log (L_{\rm H\alpha} / {\rm erg\ s^{-1} })\approx 42.5$ to $17\%$ 
at $\log (L_{\rm H\alpha} / {\rm erg\ s^{-1} }) \approx 43.3$. The most luminous \ha\ emitters include a large portion of broad-line AGNs at $z\approx 4-5$. The rising trend of AGN fraction towards the higher luminosity end is in agreement with those observed at $z\approx 0.8-2.2$ \citep{Stott2013,Sobral2016}. 
To more precisely quantify the AGN fraction at high redshift, deeper spectroscopic observations are required to resolve the broad components in fainter \ha\ emitters.

\xj{To compare the number density of BHAEs with quasars and AGNs at similar redshifts, we estimate the bolometric LF of BHAEs by converting their $L_{\rm H\alpha, broad}$ into $L_{\rm bol}$ following Equation \ref{eq:LLbol}.  The bolometric LF is shown in Figure  \ref{fig:bol_LF}.  \xj{For comparison, we also convert the broad \ha\ LF of \cite{Matthee2023} into $L_{\rm bol}$.} At the most luminous bin  ($L_{\rm bol}\approx10^{45.5
} {\rm erg ~s^{-1}}$),  our result is consistent with the extrapolation of bolometric LF of quasars at $z\sim 4$ but are $\sim 2\times$ higher than those of quasars at $z\sim 5$, despite the large uncertainty. At the faintest end ($L_{\rm bol}\approx10^{44.7
} {\rm erg ~s^{-1}}$) the number density of BHAEs is approximately $1.6\times$ that of  $z\sim 4$ quasars and $3.7\times$ that of  $z\sim 5$ quasars.  The underlying BHAE population could have a higher number density with fainter $L_{\rm H\alpha, broad}$. Our measured number density of BHAEs is generally consistent with \cite{Greene2023} and \cite{Matthee2023}, considering the uncertainties induced by dust attenuation, different selection criteria, and the cosmic variance.  
}

\xj{However, it is unclear whether it is reasonable to compare the bolometric LF of quasars with that of BHAEs.  The conversion between \ha\ and bolometric luminosity in Equation \ref{eq:LLbol} was calibrated based on the mid-infrared and optical properties of type-1 quasars at low redshifts. These high-redshift BHAEs may exhibit very different SEDs from those of typical quasars \citep[e.g.,][]{Perez2024}.  The estimated $L_{\rm bol}$ are subject to significant systematic uncertainties. The different bolometric correction recipes also introduce uncertainties \citep{Runnoe2012, Duras2020}. Multiwavelength observations and comprehensive analysis of BHAE SEDs are essential to investigate their bolometric properties further. }

\begin{figure}
    \centering
    \includegraphics[width=\columnwidth]{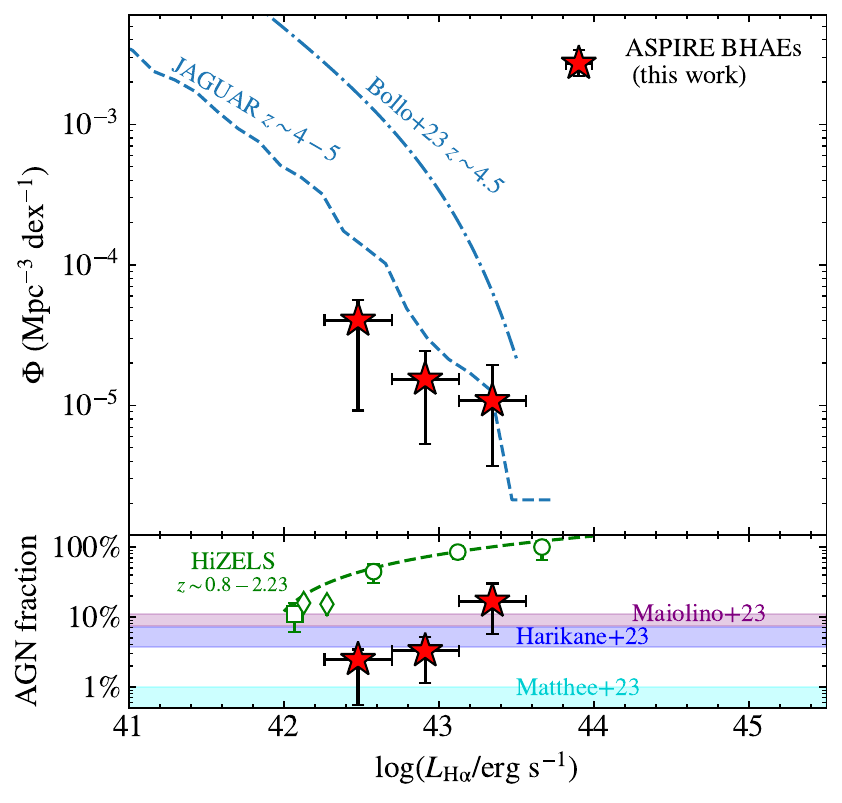}
    \caption{The \ha\ (broad + narrow) luminosity function (LF) of the BHAEs  (red stars).  The \ha\ LF of Lyman Break galaxies at $z\sim 4.5$ is presented as the blue dashed-dotted line and the prediction of \texttt{JAGUAR} mocks \citep{Williams2018} for star-forming galaxies is presented as the blue dashed line. 
    We show the fraction of BHAEs in each luminosity bin. The AGN fractions reported in \cite{Matthee2023}, \cite{Harikane2023}, and \cite{Maiolino2023} are labeled as cyan, blue, and purple-shaded regions. 
    For comparison, we show the AGN fraction from the HiZELs sample at $z\sim 0.8-2.23$: the AGN fraction based on BPT diagram \citep{Stott2013} as the green square, the AGN fraction identified using Spitzer/IRAC colors \citep{Sobral2016} as the green diamonds and the AGN fraction by broad \ha\ components \citep{Sobral2016} as the green circles. }
    \label{fig:HaLF}
\end{figure}

\begin{figure}
    \centering
    \includegraphics[width=\columnwidth]{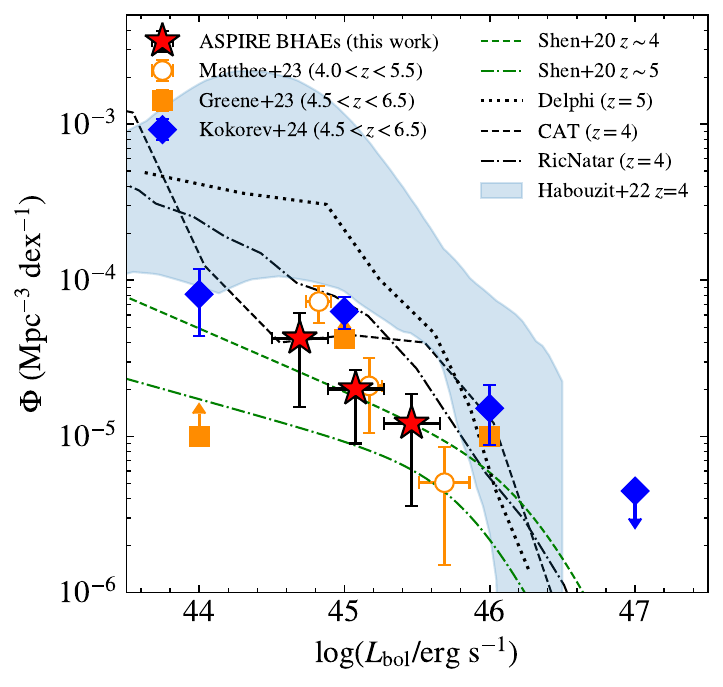}
    \caption{ The bolometric luminosity function of BHAEs (red stars). The bolometric luminosities of ASPIRE BHAEs are estimated based on the broad \ha\ luminosities. We also show the \xj{broad \ha-converted} bolometric LF of $z\approx4.5-6.5$ AGNs in \cite{Greene2023} and \xj{the $L_{5100}$-converted bolometric LF} in \cite{Kokorev2024} as the blue diamonds and orange squares, and convert the broad \ha\ LF of \cite{Matthee2023}, which were not corrected for dust attenuation either, into the bolometric LF. The bolometric LFs of quasars at $z\sim 4$ and $z\sim 5$ are shown as the gray dashed and dash-dotted lines \citep{Shen2020}.  We also present the predicted AGN bolometric LFs of \textit{Delphi} at $z=5$ \citep{Dayal2019},  
    CAT at $z=4$ \citep{Trinca2022},
    \citet{Ricarte2018} at $z=4$,
    and the range of AGN bolometric LFs at $z=4$ from different hydrodynamical cosmological simulations concluded in \cite{Habouzit2022}.
    \label{fig:bol_LF}
    }
\end{figure}

\begin{table}
 \centering
\begin{tabular}{ccccc}
    \hline
    $\log L_{\rm H\alpha}$ & $\Delta \log  L_{\rm H\alpha}$ & N & $\Phi$ ($10^{-5}$ Mpc$^{-3}$ dex$^{-1}$) & Frac$(\%)$ \\
   \hline
    42.478 & 0.433 & 6 & $4.04_{-3.12}^{+1.60}$ & $2.46_{-1.90}^{+0.97}$ \\
    
    42.911 & 0.433 & 6 & $1.53_{-1.00}^{+0.92}$ & $3.28_{-2.14}^{+1.96}$ \\
    
    43.345 & 0.433 & 4 & $1.08_{-0.71}^{+0.85}$ & $16.82_{-11.07}^{+13.30}$ \\
    \hline
    \hline
    $\log L_{\rm bol}$ & $\Delta \log  L_{\rm bol}$ & N & $\Phi$ ($10^{-5}$ Mpc$^{-3}$ dex$^{-1}$) \\
    \hline
    44.692 & 0.385 & 5 & $4.26_{-2.72}^{+1.91}$ \\
     
    45.077 & 0.385 & 7 & $2.01_{-1.11}^{+0.66}$ \\
    
    45.462 & 0.385 & 4 & $1.21_{-0.85}^{+0.66}$ \\
    \hline
    
\end{tabular}
 \caption{The \ha\ LF, BHAE fraction, and bolometric LF as shown in Figure \ref{fig:HaLF} and \ref{fig:bol_LF}. The number densities and fractions are all lower limits since we do not correct for the completeness.
 \label{tab:HaLF}
 }
\end{table}

\section{Discussion}\label{sec:discussion}

\subsection{High number density of BHAEs}\label{sec:ndensity}





Figure \ref{fig:bol_LF} suggests a high AGN number density at the lower end of the bolometric luminosity, albeit with significant potential systematic. \xlin{We note that we do not apply any reddening correction in Figure \ref{fig:bol_LF}.  If we correct $L_{\rm bol}$ using a typical value of $A_V=2$, $L_{\rm bol}$ would increase by a factor of 2.3. Applying $A_V=4$ would increase $L_{\rm bol}$ by 6.9. Consequently, the bolometric LF would shift to the bright end by 0.4-0.8 dex, increasing the excess of BHAE number density over that of type-1 AGNs.} As discussed in \cite{Greene2023}, while the number density of AGNs is larger than what would be expected from the extrapolation of the quasar luminosity function towards faint luminosities, a large number density of faint quasars is generally consistent with predictions from theoretical models, at least at $z<6.5$, while there may be some tension at the bright end for $z>6.5$. Indeed, the overprediction of the faint end of the luminosity function appeared to be common, as shown for instance in the compilation of results from large-volume simulations of \cite{Habouzit2022}. The updated AGN number densities is in much better agreement with the predicted large number of high-redshift faint AGNs. Most of these large-volume simulations, however, include simplistic predictions for black hole formation and growth, since they cannot resolve the sub-parsec scales relevant for black hole physics. 

Semi-analytical models tend to include more specific recipes for black hole formation \citep[e.g.,][]{Ricarte2018, Dayal2019, Trinca2022}. Many of these models are based on bimodal black hole seed models: ``light'' seeds with mass $<10^3$ $M_\odot$ remnants of the first stars, or ``heavy'' seeds with mass $>10^5$ $M_\odot$ formed from the collapse of supermassive stars formed under very specific conditions \citep[see][for reviews]{Inayoshi2022,Volonteri2023}. They usually find that the number density of black hole seeds decreases with the seed mass, but the ability to grow instead increases with the seed mass. Interestingly, recent simulations and theoretical models of black hole seed formation suggest a less extremely bimodal situation \citep[e.g.,][]{Regan2020,Schleicher2022}, with numerous seeds at mass $\sim 10^3$ $M_\odot$, 
more consistent with a widespread population of relatively bright AGNs presented in this paper.  

\xj{ Finally, as discussed in \S\ref{sec:LF}, the observed high-redshift BHAE population is not yet fully characterized.  The bolometric luminosity of BHAEs might not be well described by Equation \ref{eq:LLbol}. A valid comparison of BHAE number density between simulations and observations relies on thoroughly understanding BHAEs' nature, bolometric properties, selection function, and completeness through future observations.
}

\subsection{Evidence of clustering of BHAEs}


\xj{In this section we present early evidence of strong clustering of BHAEs.}

\subsubsection{A pair of BHAEs}
J1526M2050-BHAE-2 and J1526M2050-BHAE-3, both at $z=4.87$, are  1.36 arcmin apart from each other, corresponding to a \xj{projected} separation of 519 pkpc. The velocity offset between the two is $\approx 222$ km s$^{-1}$.  There might be physical connections between these two BHAEs, which could reside in the same over-density or large-scale structure.  A systematic search for galaxies at a similar redshift is needed for further detailed studies.

\subsubsection{J2232P2930-BHAE-1: an extended H$\alpha$ emitter with close neighbors}\label{sec:J2232P2930-BHAE-1}

Among the 15 ASPIRE BHAEs presented in this paper, J2232P2930-BHAE-1 \xj{at $z=4.135$} has the brightest F356W magnitude ($m_{\rm F356W}= 23.39$) and very high \ha\ luminosity ($L_{\rm H\alpha, tot}= 2\times 10^{43}$ erg s$^{-1}$).  Its F356W morphology is more extended than that in the F200W band, indicative of diffuse extended \ha\ emission. We present its F356W image, PSF-homogenized F200W contour, and the corresponding radial profiles in the left and middle panels of Figure \ref{fig:extended}. Its F115W and F200W images reveal four marginally resolved components as shown in the right panel of Figure \ref{fig:extended}. It also shows significant spatial variation in $\beta_{\rm UV}$.  The upper component \xj{($(x,y)=(0, 0.15\arcsec)$)} has a red  UV slope ($\beta_{\rm UV}\gtrsim 1$) and the left component is comparatively blue ($\beta_{\rm UV}\lesssim -0.5$).  It is not clear if there is  $\beta_{\rm UV}$ variation in the two \xj{central F200W contour peaks}, limited by the spatial resolution of F200W. The variation of $\beta_{\rm UV}$ might be caused by the inhomogeneous distribution of dust content. As a composite system with multiple components, the broad line profile of J2232P2930-BHAE-1 could be a result of blended  \ha\ emission from individual substructures, and not necessarily corresponding to an AGN broad line region. Future observations with \textit{JWST}/NIRSpec IFU are required to investigate its internal structure.

Moreover, J2232P2930-BHAE-1 has three close companion galaxies at the same redshift.  As shown in Figure \ref{fig:J2232P2930_AGN_companion}, the three companion galaxies, named Companion-1, Companion-2, and Companion-3, are 2.1 arcsec, 1.6 arcsec, and 3.85 arcsec away from J2232P2930-BHAE-1, corresponding to 14.5 pkpc, 10.8 pkpc and 26.4 pkpc at $z=4.13$. The Companion-1, Companion-2, and Companion-3 have \ha\ luminosity of $(1.7\pm 0.7) \times 10^{42}$  erg s$^{-1}$, $(1.9\pm 0.2) \times 10^{42}$  erg s$^{-1}$ and $(5.2\pm 0.5) \times 10^{41}$  erg s$^{-1}$, respectively. It implies that J2232P2930-BHAE-1 may reside in an overdense region.  There is a tentative detection of diffuse \ha\ emission between  Companion-2 and J2232P2930-BHAE-1, as shown in both the RGB image (left panel of  Figure \ref{fig:J2232P2930_AGN_companion}) and the 2D grism spectrum (middle panel of Figure \ref{fig:J2232P2930_AGN_companion}), indicating possible interaction between the two sources.  Detailed studies on the environment of J2232P2930-BHAE-1 will be performed in future works with a systematic search for \ha\ emitters at the same redshift.

  

\begin{figure*}
\includegraphics[width=.615\textwidth]{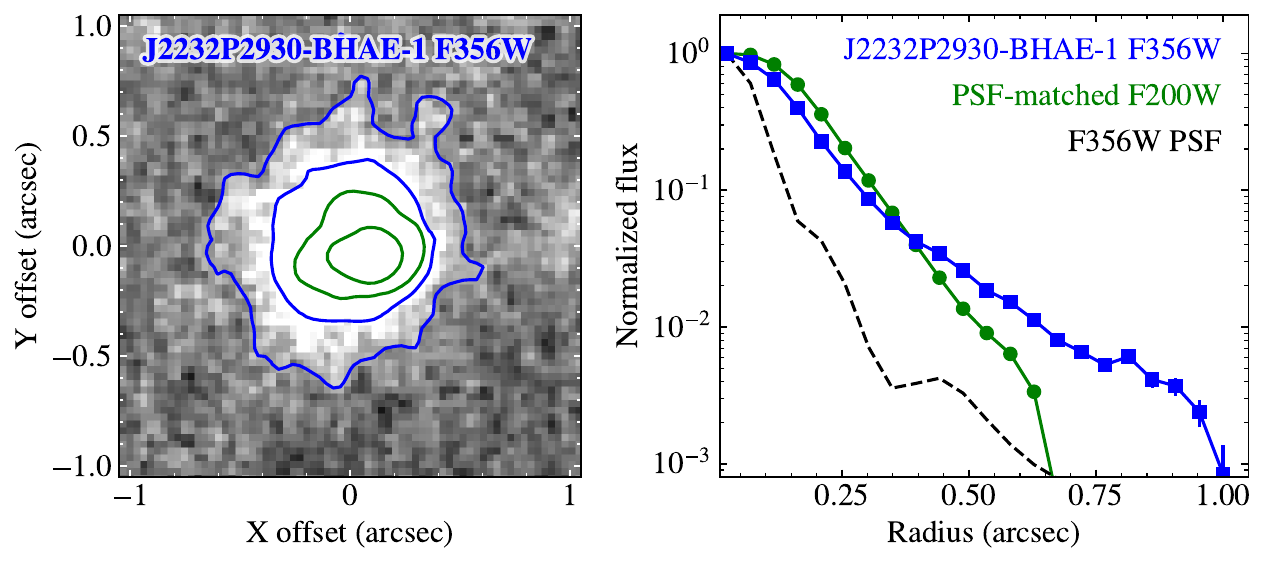}
\includegraphics[width=.385\textwidth]{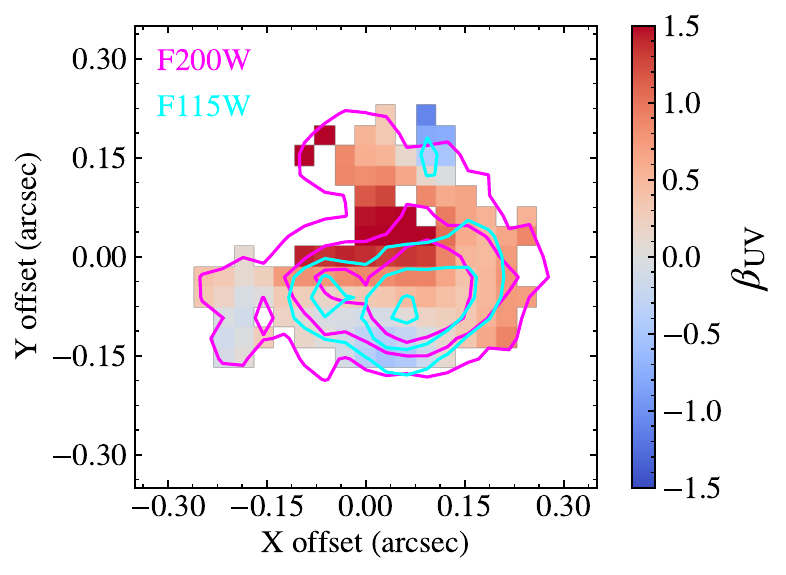}
\caption{ \textit{Left:} \textit{JWST}/NIRCam F356W image of J2232P2930-BHAE-1. The blue lines denote the 2$\sigma$ and 7$\sigma$ contours of the F356W image. The green lines are the 2$\sigma$ and 7$\sigma$ contours of the PSF-homogenized F200W image, matched to the F356W PSF.  \textit{Middle:} The surface brightness radial profile of J2232P290-BHAE-1. The blue squares show the F356W surface brightness profile of J2232P2930-BHAE-1 and the green circles show the profile of PSF-homogenized F200W. We show the radial profile of the F356W PSF as the black dashed line.  \textit{Right:} The UV slope map of J2232P2930-BHAE-1, produced using F115W and F200W images, PSF-matched to F200W. \xj{We smooth the map using a 1-pixel Gaussian kernel.} The magenta and cyan lines denote the $2\sigma$, $5\sigma$, and $10\sigma$ contours of the F115W and F200W images, respectively.
    \label{fig:extended}}        
\end{figure*}


\begin{figure*}
    \centering
    \includegraphics[width=0.95\textwidth]{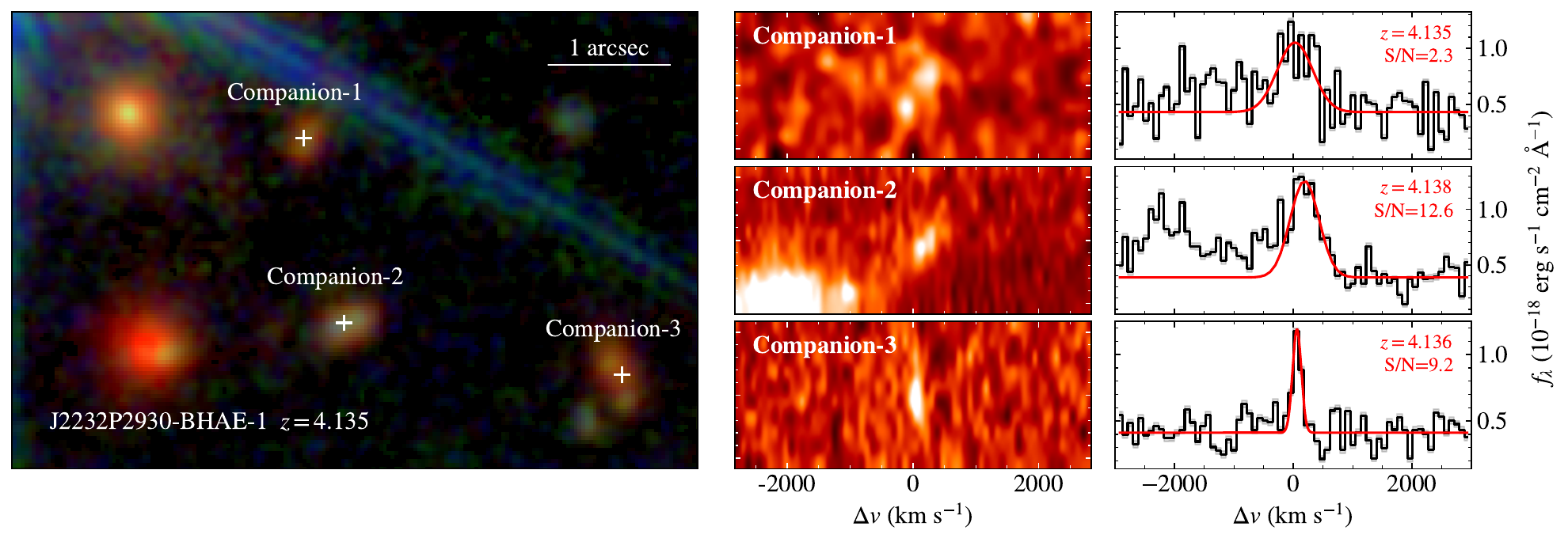}
    \caption{J2232P2930-BHAE-1 and its three neighboring galaxies. The left panel shows the RGB image composed of F115W, F200W, and F356W. We label the three companion galaxies using white plus markers. In the middle panel, we present the 2D WFSS spectra of the three galaxies, and the extracted 1D spectra are shown in the right panel. The 2D and 1D spectra are both displayed in the velocity frame relative to  J2232P2930-BHAE-1. \xj{ We can see tentative diffuse \ha\ gas between Companion-2 and J2232P2930-BHAE-1 in the RGB image and the 2D grism, suggesting possible interaction between the two sources. }}
    \label{fig:J2232P2930_AGN_companion}
\end{figure*}

\subsection{Blue-shifted \ha\ absorption}

J0923P0402-BHAE-1, J1526M2050-BHAE-1, and J1526M2050-BHAE-3 present blueshifted absorption features on top of their \ha\  emission lines. The velocity offsets between the absorption and the \ha\ emission peak for J0923P0402-BHAE-1, J1526M2050-BHAE-1, and J1526M2050-BHAE-3 are 586 km s$^{-1}$,   357 km s$^{-1}$, and 266 km s$^{-1}$, respectively.

If these features are \ha\ absorption of high-column-density gas, the fraction is high (19\% in ASPIRE BHAEs, 10\% in the sample of \citealt{Matthee2023}, 15\% in the sample of \citealt{Kocevski2024}) compared to low-redshift type-1 AGNs \citep[$<0.1\%$, e.g.,][]{Aoki2006, Zhang2018}, as shown in Figure \ref{fig:Balmer_absorption_fraction}. The prevalence of \ha\ absorption, if confirmed, suggests that these high-redshift BHAEs could differ significantly from low-redshift type-1 AGNs. High-redshift BHAEs may contain abundant outflowing dense gas, which could be related to either BH activity or strong galactic feedback in the host galaxies. 


\begin{figure}[h!]
    \centering
    \includegraphics[width=\columnwidth]{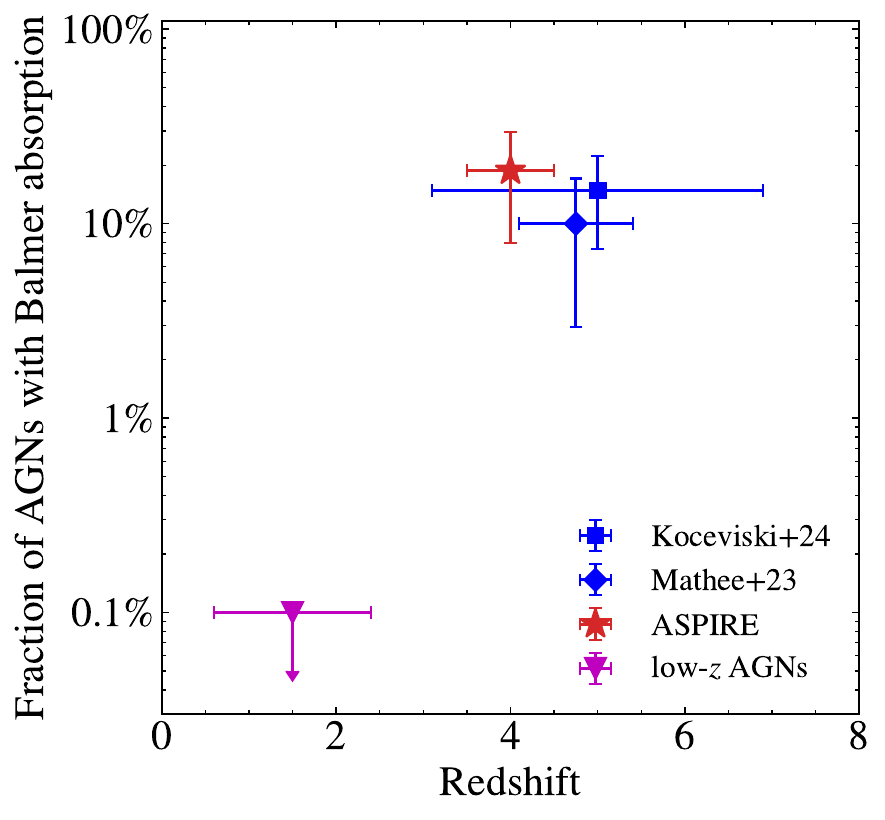}
    \caption{ The fraction of AGNs with Balmer absorption across the cosmic time. The errorbars in the redshift axis indicate the redshift range of the AGN sample, while the uncertainties in the fraction on the $y$-axis are calculated assuming a Poisson distribution.}
    \label{fig:Balmer_absorption_fraction}
\end{figure}

\subsection{Comparison with literature LRDs and selection effect}
\begin{figure*}
    \centering
    \includegraphics[width=0.305\textwidth]{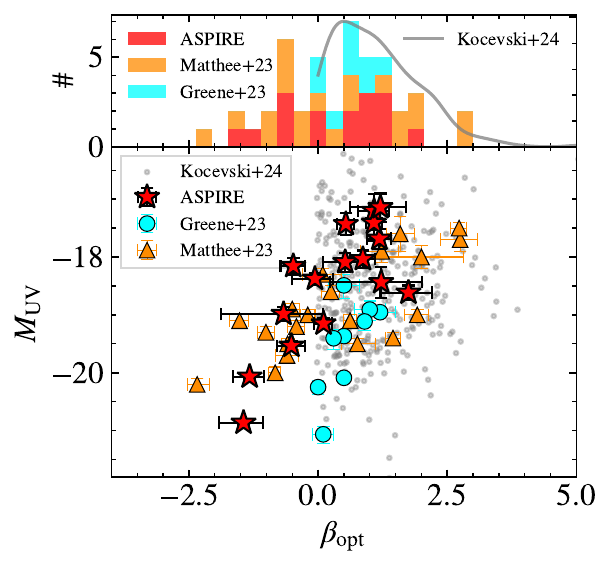}
     \includegraphics[width=0.305\textwidth]{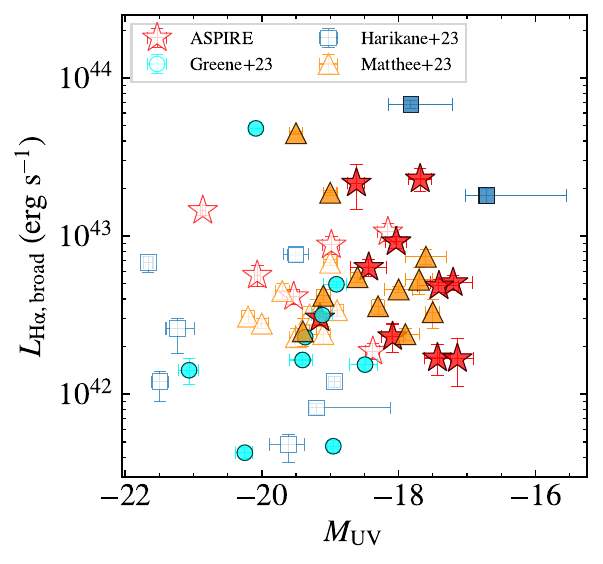}
    \includegraphics[width=0.365\textwidth]{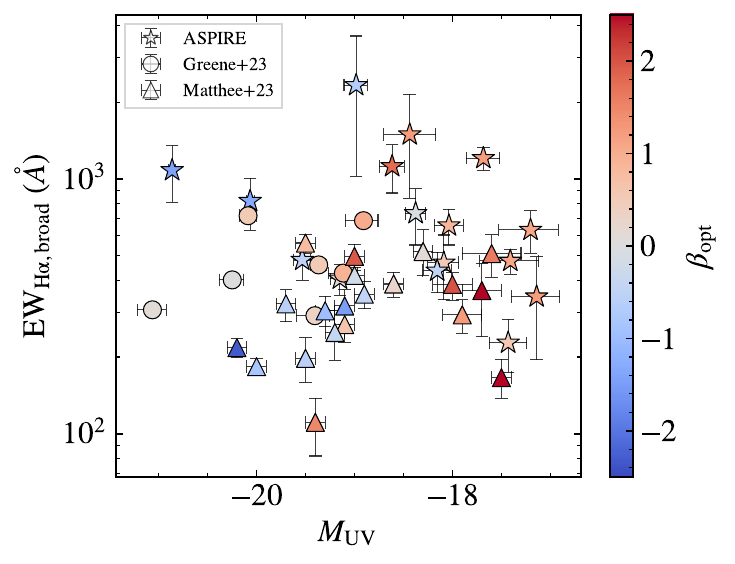}
    \caption{Left panel: the \betaopt\ versus $M_{\rm UV}$ distribution of JWST broad-line and photometrically selected LRDs. The broad-line selected samples include the ASPIRE BHAEs (red) and the BHAEs in \cite{Matthee2023}  (orange). The photometrically selected samples include spectroscopically confirmed BHAEs in \cite{Greene2023} (cyan) and LRDs in \cite{Kocevski2024} (gray). We show the density distribution of \betaopt\ for the \citealt{Kocevski2024} sample. Middle panel: the broad \ha\ luminosity versus $M_{\rm UV}$. The filled markers denote the BHAEs with reddened optical continua (\betaopt$>$0) while white ones have \betaopt$<0$. Right panel:  the rest-frame equivalent width of the broad \ha\ versus $M_{\rm UV}$, color-coded by  \betaopt.}
    \label{fig:MUV_beta_opt}
\end{figure*}

\xlin{
We compare photometrically and broad-line selected samples in the literature in Figure  \ref{fig:MUV_beta_opt}.  The broad-line selected samples (e.g., ASPIRE, \citealt{Matthee2023}, and \citealt{Harikane2023}), which imposes no constraints on the SED shapes,  contain BHAEs with \betaopt$\lesssim0$, as illustrated in the left panel of Figure \ref{fig:MUV_beta_opt}. On the other hand,  the samples photometrically selected based on V-shaped SEDs (\citealt{Greene2023} and \citealt{Kocevski2024}) do not include such objects.  Meanwhile, the broad-line selected BHAEs lack UV-luminous but red objects (with $M_{\rm UV} \lesssim -19.5$ and \betaopt$>0$).

The photometric selection requiring V-shaped SEDs is biased towards sources with reddened rest-frame optical spectra and against those with relatively blue optical continua (\betaopt$<0$).  The flux-limited broad-line selection fails to detect objects with weak broad \ha\ flux, as their \ha\ broad wings could be overwhelmed by background noise. As shown in the middle panel of Figure \ref{fig:MUV_beta_opt}, the UV-luminous and red sources in \cite{Greene2023} exhibit broad \ha\ luminosity lower than those of most \textit{JWST}/WFSS-selected BHAEs. Their broad \ha\ components may not be reliably identified with the shallow grism data. We note that among the NIRSpec-selected sample from \cite{Harikane2023}, the two sources with reddened rest-frame optical continua also have the faintest  $M_{\rm UV}$ ($M_{\rm UV}>-18$). Since their broad-line selection relies on archive data, the prior source selection for the NIRSpec MSA is crucial but can be intricate. 

The distribution of grism-selected BHAEs on the \betaopt$-M_{\rm UV}$ and $M_{\rm UV}-{\rm EW_{H\alpha, broad}}$ planes  (the left and right panels of Figure \ref{fig:MUV_beta_opt}) is attributed to the flux-limited selection effect.  The absence of UV-faint and optical-blue objects ($M_{\rm UV} \gtrsim -18$ and \betaopt$<0$) can be attributed to their weak broad \ha\ wings falling below the detection limit, as they might have lower-mass BHs and host galaxies following the co-evolution of the BH-galaxy. Although the UV-luminous and optically red BHAEs ($M_{\rm UV} > -19$ and \betaopt$>0$) missing from grism-selected sample all have low $L_{\rm H\alpha, broad}$, similar objects with high $L_{\rm H\alpha, broad}$ might exist but are rare due to the rapid decline towards the bright end of the LFs.  On the other hand, it remains unclear whether BHAEs with \betaopt$>0$ and \betaopt$<0$ belong to the same population. A plausible interpretation is that \betaopt$<0$ BHAEs may closely resemble type-1 quasars. It is also possible that BHAEs with \betaopt$<0$ are counterparts of those with \betaopt$>0$, but with less obscured AGN components. In this case, if the UV emission originates from the host galaxies, the stellar light from UV-luminous star-forming galaxies may contribute significantly to the rest-frame optical spectra, potentially leading to a bluer \betaopt.

 We notice that the two optically red BHAEs identified in \cite{Harikane2023} are both compact and point-source-like, in contrast to the other objects they selected, which are either extended or have bright companions.  The depth of ASPIRE images does not allow us to resolve the morphology of individual BHAEs. Instead, we examine the BHAEs in \cite{Matthee2023}. Among the nine BHAEs with \betaopt$<0$ in \cite{Matthee2023}, seven have extended F200W (or F182M + F210M) morphologies, and only two are compact and point-source-like. Among the 11 \betaopt$>0$ BHAEs, four show significant extended F200W morphologies and the remaining seven are compact. The `compactness rate' is 63\% for \betaopt$>0$ BHAEs compared to 22\% for \betaopt$<0$ BHAEs. This can also be due to a selection effect. As discussed above, the grism-selected BHAE sample misses UV-luminous but optically red objects, and as shown in the left panel of Figure \ref{fig:MUV_beta_opt}, the grism-selected \betaopt$>0$ BHAEs are all UV faint. Their UV faintness makes it challenging to resolve the extended components in short-wavelength images. To conclusively determine if there is any potential correlation between optical redness and morphologies, a larger and unbiased sample is essential.
 
In summary, different selection criteria introduce various biases that must be carefully considered to understand the nature of this newly discovered BHAE population. A systematic classification is essential for understanding their properties. To explore a broader range of samples in the future, a more inclusive selection criterion that considers both AGN and galaxy contributions in the rest-frame optical band is necessary.

}

\section{Summary}\label{sec:summary}

We report the discovery of 16 broad \ha\ emitters (BHAEs) at $z=4-5$ selected from the JWST ASPIRE program. These BHAEs are primarily selected due to their compactness and redness. Their broad \ha\ emission lines are identified utilizing the \textit{JWST}/NIRCam WFSS of F356W. Our main results are summarized below.

\begin{itemize}
    \item [(1)] The ASPIRE BHAEs exhibit broad \ha\ emission components with FWHM (FWHM$_{\rm broad}$) ranging from 1000 to 3000 km s$^{-1}$ and luminosity ($L_{\rm H\alpha, broad}$) from $10^{42}$ to $\gtrsim 10^{43}$ erg s$^{-1}$.  The fraction of \ha\ flux in the broad components ($L_{\rm H\alpha, broad}/L_{\rm H\alpha, total}$) spans a wide range from $\sim 0.4$ up to 1 with a median value of 0.7. These BHAEs are distributed similarly in the ${\rm FWHM_{ broad}}-L_{\rm H\alpha, broad}$ diagram compared to low-redshift ($z<0.6$) SDSS quasars. On the other hand, BHAEs have lower $L_{\rm H\alpha, broad}/L_{\rm H\alpha, total}$ than most of the SDSS quasars, suggesting larger contributions from the host galaxies or narrow line regions to the total \ha\ luminosity.

    \item[(2)] The black hole masses (\MBH) of ASPIRE BHAEs range from $\sim 10^{7}$ to $\sim 10^{8}$ $M_\odot$, comparable to those of BHs in low-redshift SDSS quasars.  The \ha\ converted bolometric luminosity (\Lbol) suggests that these BHAEs are accreting at eddington ratios of 0.07-0.47, with a median value of 0.17.

    \item[(3)]  In general, the ASPIRE BHAEs have UV continuum slopes (\betauv) bluer than the optical continuum slopes (\betaopt).  In contrast to \cite{Matthee2023}, we do not see significant correlations between \betaopt\ and $L_{\rm H\alpha, broad}/L_{\rm H\alpha, total}$ or \MBH. 

    \item[(4)] The stacked images of ASPIRE BHAE show extended components in F200W. The F200W stack extends to $\sim 0.4$ arcsec, corresponding to 2.6 physical kpc at $z\sim 4.5$. The extended morphology in the F115W and F356W are tentative, and limited by the shallow depth of observations.  The extended structure suggests the presence of host galaxies.

    \item[(5)] The \ha\ LF of ASPIRE BHAEs suggests that the AGN fraction increases towards higher \ha\ luminosities, consistent with low-redshift \ha\ emitters. The BHAE fraction spans from $2\%$ at total \ha\ luminosity of $\log(L_{\rm H\alpha}/{\rm erg\ s^{-1}})\approx 42.5$ to $17\%$  at $\log(L_{\rm H\alpha}/{\rm erg\ s^{-1}})\approx 43.3$. 

    \item[(6)] We find a pair of BHAEs at the same redshift ($z\approx 4.87$) with a separation of 519 physical kpc. \xj{Another isolated BHAEs,} J2232P2930-BHAE-1, as a composite system with bright extended \ha\ emission and significant \betauv\ variation, resides in an overdense region with three close companion galaxies. These, for the first time, provide \xj{tentative} evidence for the strong clustering of BHAEs.    
    
    \item[(7)] We find three BHAEs with blueshifted \ha\ absorption, \xj{potentially} indicating the presence of outflowing dense gas. The prevalence of \ha\ absorption (19\% in ASPIRE BHAEs) might be a unique feature of this high-redshift AGN population.

    \item[(8)]  \xlin{We compare the broad-line and photometrically selected BHAE samples from the literature. The photometrically selected sample, based on V-shaped SED, does not contain optically blue ($\beta_{\rm opt}<0$) objects, while the flux-limited grism-selected sample lacks UV-luminous but optically red ($M_{\rm UV} \lesssim -19.5$ and $\beta_{\rm opt}>0$) objects. The optically red BHAEs tend to be more compact in rest-frame UV bands. However, these differences can be explained by the selection effect. A more inclusive selection method is needed to study a broader range of BHAEs.}
\end{itemize}

The ASPIRE BHAE sample with spectroscopically identified broad \ha\ emission provides a good database for further studies on low-luminosity AGNs in the early Universe. To further investigate the nature of high-redshift low-luminosity AGNs and their roles on the evolutionary tracks of early BH assembly,  future follow-ups in both the restframe UV and infrared wavelengths are required.  Deep spectroscopic and IFU observations covering the UV continua and emission lines can provide constraints on the origin of the extended blue components.  Observations in the infrared band with \textit{JWST}/MIRI are crucial to \xj{determine the full SED shapes} and to understand the roles of AGNs and starbursts in these broad-line selected samples.

\acknowledgments

X.L. and Z.C. are supported by the National Key R\&D Program of China (grant no.~2023YFA1605600), the National Science Foundation of China (grant no.~12073014), the science research grants from the China Manned Space Project with No.~CMS-CSST-2021-A05, and Tsinghua University Initiative Scientific Research Program (No.~20223080023). F.W. and X. F. acknowledge support from NSF Grant AST-2308258. F.S. acknowledges JWST/NIRCam contract to the University of Arizona NAS5-02015. JBC acknowledges funding from the JWST Arizona/Steward Postdoc in Early Galaxies and Reionization (JASPER) Scholar contract at the University of Arizona.  AL acknowledges support by the PRIN MUR ``2022935STW". MT acknowledges support from the NWO grant 016.VIDI.189.162 (``ODIN"). JTS is supported by the Deutsche Forschungsgemeinschaft (DFG, German Research Foundation) - Project number 518006966.

This work is based on observations made with the NASA/ESA/CSA James Webb Space Telescope. The data were obtained from the Mikulski Archive for Space Telescopes at the Space Telescope Science Institute, which is operated by the Association of Universities for Research in Astronomy, Inc., under NASA contract NAS 5-03127 for JWST. These observations are associated with program \#2078.  The specific observations analyzed can be accessed via \dataset[10.17909/vt74-kd84]{https://doi.org/10.17909/vt74-kd84}. Support for this program was given through a grant from the Space Telescope Science Institute, which is operated by the Association of Universities for Research in Astronomy, Inc., under NASA contract NAS 5-03127.

\facility{JWST}
\software{JWST calibration pipeline \citep[v1.10.2][]{Bushouse2022}  SciPy \citep{scipy} SourceExtractor++ \citep{SEpp}
}

\bibliography{main}{}
\bibliographystyle{aasjournal}

\end{document}